# Constructing BERT Models: How Team Dynamics and Focus Shape AI Model Impact

**Likun Cao** (Department of Sociology, Purdue University; Knowledge Lab, The University of Chicago; 0000-0001-5234-3855)
**Kai Li** (University of Tennessee, Knoxville, Knoxville, TN, USA; 0000-0002-7264-365X)

Corresponding author: Kai Li, kli16@utk.edu

Abstract

The rapid evolution of AI technologies, exemplified by BERT-family models, has transformed scientific research, yet little is known about their production and recognition dynamics in the scientific system. This study investigates the development and impact of BERT-family models, focusing on team size, topic specialization, and citation patterns behind the models. Using a dataset of 4,208 BERT-related papers from the Papers with Code (PWC) dataset, we analyze how the BERT-family models evolve across methodological generations and how the newness of models is correlated with their production and recognition. Our findings reveal that newer BERT models are developed by larger, more experienced, and institutionally diverse teams, reflecting the increasing complexity of AI research. Additionally, these models exhibit greater topical specialization, targeting niche applications, which aligns with broader trends in scientific specialization. However, newer models receive fewer citations, particularly over the long term, suggesting a "first-mover advantage," where early models like BERT garner disproportionate recognition. These insights highlight the need for equitable evaluation frameworks that value both foundational and incremental innovations. This study underscores the evolving interplay between collaboration, specialization, and recognition in AI research.

**Keywords:** Knowledge production; Artificial Intelligence; BERT; Citation

1 Introduction

Science and technology have long shared a co-evolutionary relationship, with technological advancements providing the instruments and capacities that expand the boundaries of scientific inquiry (Brooks, 1994; Rip, 1992). In the 21st century, as science becomes increasingly data-driven and computationally intensive, technology has emerged as a dominant force in scientific development. This growing asymmetry is evident in concepts like *AI4Science* (Stevens et al., 2020; Xie et al., 2024) and the proliferation of fields prefixed with "computational," such as *computational biology* and *computational social science* (Lazer et al., 2009; Nobel, 2002). As the technological landscape evolves rapidly, understanding the development and application of emerging tools—especially AI technologies—has become critical for comprehending the scientific system and its interactions with technological advancements (Gao & Wang, 2024; Wang et al., 2023).

One of the most transformative developments in AI over the past decade has been the emergence of BERT (Bidirectional Encoder Representations from Transformers), a pre-trained language model that revolutionized natural language processing (NLP) by introducing deep bidirectional context understanding (Devlin et al., 2019). Unlike earlier models like ELMo (Embeddings from Language Models) and LSTM (Long Short-Term Memory), which process text sequentially, BERT employs a transformer-based architecture that simultaneously considers both left and right contexts, resulting in richer language representations. This innovation positioned BERT as a foundational model in modern NLP and AI research, with widespread adoption in scientific inquiry (Beltagy et al., 2019).

Following its release, BERT sparked a wave of derivative models, expanding into what is now known as the BERT family. These variants introduced functional improvements, efficiency gains, and domain-specific adaptations. Notable examples include RoBERTa (Liu et al., 2019), ALBERT (Lan et al., 2019), and BioBERT (Lee et al., 2020). The proliferation of these models underscores BERT's foundational impact on the field of AI and highlights the dynamic evolution of AI tools through collaborative scientific efforts.

Despite the significance of BERT-family models in AI research and the numerous variants developed since the original BERT release, little is known about how their development has evolved over time and how they are recognized within the scientific community. This study focuses on the inputs of these BERT-family models (team size and seniority) and their outputs (impact, measured by citations) to examine the conditions of production and recognition throughout the lifecycle of these models.

This study investigates three key questions:

1. Do newer BERT models require larger and more diverse research teams for their development?
2. Do newer BERT models focus on narrower research topics?
3. Do these newer models receive comparable recognition in terms of citations?

By exploring these dynamics, we aim to provide insights into the evolving structure of AI development and research and its broader implications for the scientific community's recognition of technological advancements in science. This, we believe, offers a novel perspective on the landscape of AI development and its application in scientific research.

Based on our research questions, we developed the following three central hypotheses:

**Hypothesis 1: Newer BERT-family models are built by larger and more diverse teams.** This hypothesis is grounded in the widely accepted view that science is increasingly conducted through large-scale collaborations (Cronin, 2021; Larivière et al., 2015; Wuchty et al., 2007). The growing complexity and specialization of AI research necessitate diverse expertise in areas such as machine learning, data engineering, and domain-specific knowledge (e.g., biomedical or legal applications). Moreover, the global nature of AI research fosters international collaborations, further diversifying team compositions (AlShebli et al., 2024; Hu et al., 2020). Specifically, we propose two sub-hypotheses regarding the levels of authors and institutions based on these observations:

**Hypothesis 1A: Newer BERT-family models involve more collaborators.** As AI models become more sophisticated, their development requires integrating diverse skill sets from individual researchers, such as algorithm design, large-scale data processing, and fine-tuning for specific applications. In this sub-hypothesis, we use the number of collaborators, a basic measurement for the size and complexity of scientific collaboration (Newman, 2001; Sonnenwald, 2007).

**Hypothesis 1B**: **Newer BERT-family models involve more institutions.** Similarly, we hypothesize that AI models increasingly require inputs from different institutions, another important indicator used in scientometrics (Gazni et al., 2012).

**Hypothesis 2: Publications on newer BERT-family models have narrower conceptual scopes.** This hypothesis is informed by the inevitable specialization of science: where newer and smaller niches are created in science over time (Ben-David & Collin, 1966; Price, 1969; Wray, 2005). Furthermore, the hypothesis is also underlined by the observation that incremental developments are often tailored to specific applications and real-world scenarios (Palmer & Brookes, 2002). Based on this, we propose that newer models are associated with narrower topical coverage, as they tend to target more specific use cases.

**Hypothesis 3: Newer BERT-family models receive fewer citations.** This hypothesis draws on the conceptualization of citations as indicators of attention received by citable objects (Parolo et al., 2015; Petersen et al., 2014). According to this perspective, publications that are more accessible (e.g., open access or publicly available datasets) or widely disseminated (for example, through social media) are more likely to attract citations (Langham-Putrow et al., 2021; Piwowar & Vision, 2013; Wang et al., 2015). Moreover, studies introducing novel ideas tend to receive greater attention and citations (Bornmann & Daniel, 2008; Foster et al., 2015; Newman, 2009). In this study, we hypothesize that newer publications receive fewer citations over time because they are relatively more "derivative" than the original BERT models.

3 Data and Methods

**3.1 Datasets**

In this research, we used the following two datasets: research articles from Papers with Code (PWC) and OpenAlex. PWC is a comprehensive, open-source repository that connects academic papers with their corresponding code implementations (Kardas et al., 2020). It primarily covers machine learning and artificial intelligence research, providing structured metadata such as paper titles, abstracts, publication venues, research tasks, methods, and performance benchmarks. This data source was selected because it offers a relatively clear and well-defined boundary of "AI publications". We mapped all publications indexed in PWC to Unpaywall (https://unpaywall.org/) to retrieve paper DOIs, and used all retrieved publication DOIs to collect metadata from OpenAlex, one of the most frequently used research databases in science of science and scientometric research (Priem et al., 2022).

The PWC-OpenAlex dataset contains 477,718 papers in total, of which 9,485 mentions the phrase "BERT" in abstracts. This research focuses on the latter sample. Within this sample, some instances of "BERT" do not refer to the BERT model, but instead have alternative meanings, such as mathematical terms and human names. To distinguish BERT-related terms (e.g., RoBERTa) from irrelevant nouns (e.g., Hilbert Space), and accurately identify references to the true BERT concepts, we employed a deep learning model to conduct the classification.

We first fine-tuned a BERT-base-uncased model on all Papers with Code abstracts. We then generated contextual word embeddings for each BERT-related concept in each abstract. Contextual word embeddings allow us to generate different vectors for the same word depending on its usage in different contexts. For example, in the sentence "He sat on the river bank and watched the water flow" and in another sentence "She went to the bank to deposit her paycheck," the word "bank" would have different embedding vectors, with the first closer to concepts like water and nature, while the second better linked to financial terms. Since our data is primarily focused on the topic of BERT, most instances of 'BERT' are expected to refer to BERT models. A concept is therefore classified as true BERT if its cosine similarity to the average embedding of 'BERT' exceeds a predefined threshold. Based on human evaluation, we set this threshold at 0.3. To examine the accuracy of our classifier, we validated it using a human-labeled validation set (N=500) (see Appendix A). The model achieves an accuracy of 0.97 and an F1-score of 0.95, confirming its effectiveness in distinguishing BERT-related concepts. In total, 4,208 papers have been identified as related to the "BERT" algorithm, rather than to human names or mathematical terms.

We also conducted robustness checks using an alternative sample inclusion criterion. As shown in the Appendix B, the results remain consistent.

## 3.2 Variables

***Independent Variable:*** *Generation of BERT Models.* For computer science studies, the generation of models can be intertwined with objective publication time. To disentangle these two presentations of "newness" of a BERT model, we quantified the methodological generation of BERT papers using citation relationships, a widely used proxy for knowledge flow (Alcácer & Gittelman, 2006; Kang et al., 2024; Price, 1969).

We began by constructing a concept network of BERT-related terms based on citation links (e.g., BERT supports the downstream concept of BioBERT; therefore, an arc is drawn from BERT to BioBERT). Next, we applied the graph hierarchy algorithm (Moutsinas et al., 2021) to estimate a generation score for each BERT-related concept. Finally, we determined the methodological generation of a paper by calculating the average generation score of all BERT-related concepts appearing in its abstract. This procedure is illustrated in Figure 1, Panel A, as a conceptual figure.

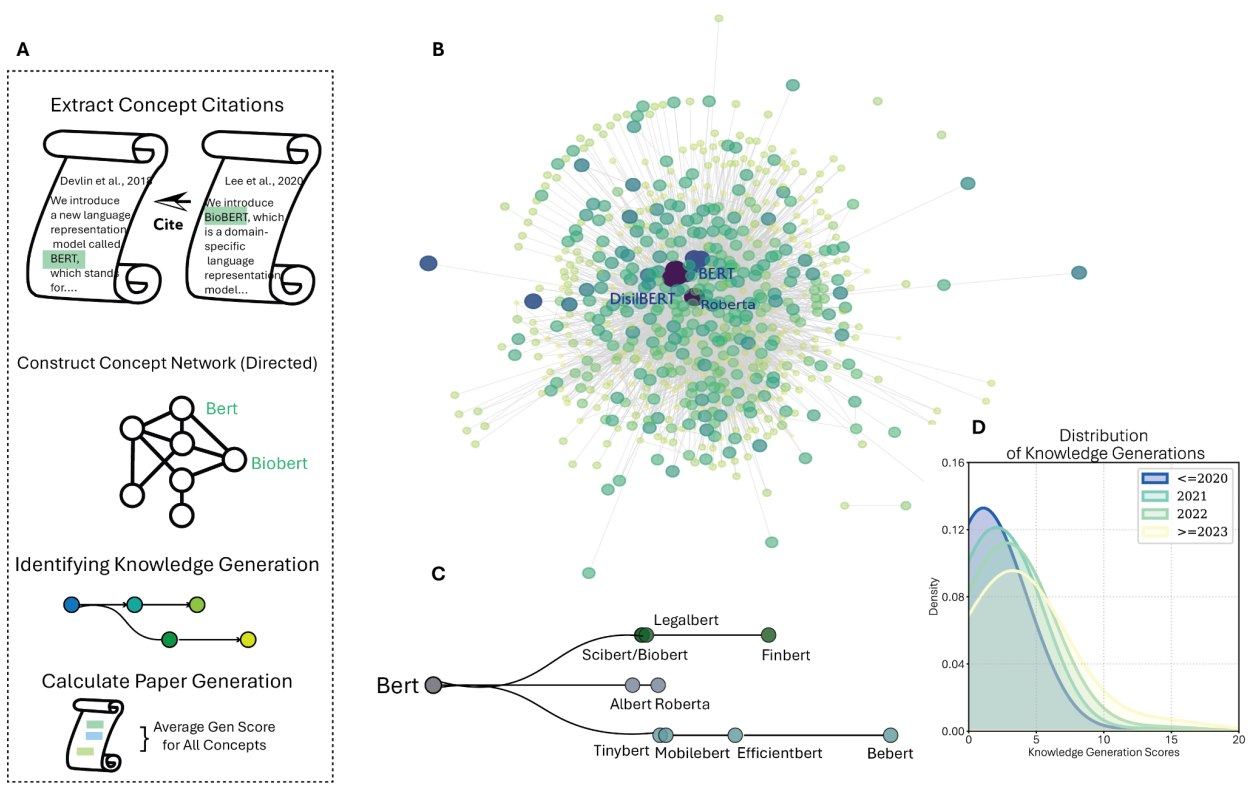

Figure 1. Generation Scores of BERT Models: Example and Distribution

Panel B displays the citation network among BERT-related terms, using a force-directed layout (with sfdp algorithm in NetworkX package), while Panel C presents the generation scores of several well-known BERT-family models along a timeline. These figures suggest that knowledge diffuses from earlier concepts, such as BERT, RoBERTa, and DistilBERT, to downstream concepts like FinBERT and EfficientBERT. Panel D in Figure 1 shows the distribution of generation scores for papers published in different years.

While the generation score is positively correlated with the objective publishing time, the two measures capture different aspects of the knowledge life cycle. When both variables are included in the model, the generation score reflects the position of the focal BERT-family model within the knowledge trajectory—that is, whether it serves as a precursor supporting other models, net of the effect of chronological publication order. We argue that the generation effect is theoretically important, as objective publication time is often entangled with period effects and broader historical trends, whereas generation score captures the underlying dynamics we aim to identify: variations in research inputs and outputs shaped by the internal mechanisms of knowledge production and the structure of intellectual dependencies within the knowledge system.

***Dependent Variable Set 1:*** *Research Input.* We measured research inputs using three sets of variables, each capturing the amount of research resources involved in a project from the perspectives of authors and institutions. **For authors**, we considered: (1) The number of authors involved in each work; (2) The average career age of authors, calculated as the number of years since their first publication; (3) The average cumulative citations previously received by each author in the publication year of the focal project (log-transformed to reduce over-dispersion). These variables reflect both the experience and expertise of authors, as well as the combined capabilities. **For institutions**, we measured: (1) The total number of unique institutions involved in each work; (2) The average number of papers produced by each institution in the publication year of the focal project (log-transformed); (3) The dispersion of authors across institutions, measured using 1 - the Herfindahl index (based on author distribution across institutions). These variables capture both the research capacity of each institution and the extent of cross-institutional collaboration. For **international collaborations**, we considered: (1) the number of countries involved in the research project, based on the locations of the institutions; and (2) the dispersion of authors across countries, measured using 1 - the Herfindahl index (based on author distribution across countries). These variables capture the extent to which a given project involves international collaboration.

***Dependent Variable Set 2:*** *Research Concepts Breadth.* We used one of the simplest and most intuitive measures—the number of concepts—to capture the breadth of knowledge of each publication. OpenAlex concepts are derived from the Microsoft Academic Graph, enabling comparison with existing research and ensuring methodological continuity. This taxonomy organizes academic knowledge into three levels. Level 0 represents broad disciplines such as *Biology, Computer Science,* and *Sociology*, while Levels 1 and 2 provide more specific categories, such as *Machine Learning* (Level 1) and *Neural Networks* (Level 2). The number of research concepts generally reflects the extent to which a paper engages with multiple disciplines.

In this study, we included all concepts provided by the OpenAlex API without any threshold. In the Appendix, we further applied different cut-off scores (0.05, 0.1, and 0.2) to determine the minimum level of topic relevance required for inclusion. We re-estimated the models using these thresholds and reported the results in Appendix Table 22, which show that the threshold does not have a strong impact on the outcome.

***Dependent Variable Set 3:*** *Research Impact.* Given the fast iteration of computer science, and the short history of BERT research (with first paper published in 2018), here we measured short-term to long-term impact as the number of citations a paper receives within one, two and three years. They are computed as follows:

$$I_{1y} = \sum_{i=0}^{1} C_i, \quad I_{2y} = \sum_{i=0}^{2} C_i, \quad I_{3y} = \sum_{i=0}^{3} C_i.$$

where $C_i$ is the number of citations each paper receives at the i-th year since its publication. Higher citations mean the paper has higher research impact.

***Control Variables.*** For all models, we controlled for the following factors: (1) Publication Venue Features: Conference (1 = Yes, 0 = No), Core Collection (whether the publication channel is identified as a "core source" in OpenAlex dataset[1]) (1 = Yes, 0 = No), and Impact Factor of the journal or conference. (2) Authors' Affiliation and Experience: Any Industrial Affiliation (1 = Yes, 0 = No), and the average number of BERT-related papers the authors had published prior to the focal publication. (3) Main Domains: we included knowledge domain fixed effects to account for differences across broad academic fields. The domain classifications are based on the Level 0 field index in the OpenAlex data, which includes 19 unique fields in total. (4) Publication Time: we also incorporated publication time fixed effects to control for temporal variations in publication trends.

For the models on research input, we controlled for the one-year forward citation count as a proxy for research quality, and the number of Level 2 concepts. For the models on knowledge breadth, we controlled for the one-year forward citation count, the number of authors, and the number of institutions. For the models on research impact, we controlled for the number of authors, the number of institutions, and the number of Level 2 concepts.

### 3.3 Model Specification

To estimate the effect of methodological generations on research resources and impact, we performed multivariate regressions, controlling for key confounders such as publication venue characteristics, subfields, and publication years. Since our dependent variables are primarily count data (e.g., citations), we used negative binomial regression models. For the continuous dependent variables, such as average career age, we used OLS regressions.

The full models are specified as below:

$$ResearchInput_i = \beta_1 \times Generation_i + \beta_2 \times Controls_i + F_i + Period_i + \epsilon_i \tag{1}$$
$$KnowledgeBreadth_i = \beta_1 \times Generation_i + \beta_2 \times Controls_i + F_i + Period_i + \epsilon_i \tag{2}$$
$$ResearchImpact_i = \beta_1 \times Generation_i + \beta_2 \times Controls_i + F_i + Period_i + \epsilon_i \tag{3}$$

Where $ResearchInput_i$ represents the research resources involved in a paper *i,* including variables related to authors and institutions; $KnowledgeBreadth_i$ represents the number of concepts assigned to the focal paper *i*; $ResearchImpact_i$ represents the number of citations a paper *i* receives within a given period (one, two, or three years); and $Generation_i$ represents the paper *i*'s

---

generation score. In addition, we included a set of features specific to paper i as control variables ($Controls_i$) and incorporated domain fixed effects ($F_i$) in the models.

To ensure the robustness of our results, we conducted stepwise regressions with forward selection to observe how the coefficients change as variables are added to the models.

We also conducted additional robustness tests to ensure the reliability of our findings, as reported in the Appendix B. All analyses are conducted using Stata 18.0.

## 3.4 Descriptive Statistics

The mean and standard deviation of all variables are reported in Table 1. The correlation matrix for all variables is presented in Table 2. While variables within the same category (e.g., one-, two-, and three-year citations) are correlated, most variables show only weak correlations.

Table 1. Statistical Descriptions of Research Input and Research Impact

| | BERT-Related Papers (N=4208) | | | |
|---|---|---|---|---|
| | Mean | SD | Min | Max |
| Number of Authors | 4.196 | 2.113 | 1.000 | 27.000 |
| Average Career Age | 6.737 | 2.318 | 0.000 | 12.000 |
| Log(Cumulative Citations) | 6.512 | 1.899 | 0.000 | 11.149 |
| Number of Institutions | 1.787 | 1.155 | 1.000 | 15.000 |
| Log(All Paper Counts) | 8.631 | 3.742 | 0.000 | 12.849 |
| Institution Dispersion (1-HHI) | 0.279 | 0.080 | 0.000 | 0.893 |
| Number of Countries | 1.298 | 0.352 | 1.000 | 7.000 |
| Country Dispersion (1-HHI) | 0.123 | 0.043 | 0.000 | 0.820 |
| Number of Level 0 Concepts | 3.197 | 1.256 | 1.000 | 8.000 |
| Number of Level 1 Concepts | 5.737 | 1.970 | 0.000 | 13.000 |
| Number of Level 2 Concepts | 5.420 | 2.476 | 0.000 | 14.000 |
| One Year Citations | 3.330 | 4.031 | 0.000 | 25.000 |
| Two Year Citations | 6.368 | 6.095 | 0.000 | 25.000 |
| Three Year Citations | 8.579 | 7.736 | 0.000 | 25.000 |
| Paper Generation Score | 2.350 | 2.125 | 0.000 | 14.880 |
| Publication Year | 2021.477 | 1.368 | 2018.000 | 2024.000 |
| Venue: Conference | 0.755 | 0.430 | 0.000 | 1.000 |
| Venue: Core Collection | 0.177 | 0.382 | 0.000 | 1.000 |
| Venue: Impact Factor | 1.977 | 3.365 | 0.000 | 27.041 |
| Any Industrial Affiliation (Dummy) | 0.146 | 0.353 | 0.000 | 1.000 |
| Previous BERT Publications | 0.400 | 0.816 | 0.000 | 8.000 |

Table 2. Correlation Matrix for All Variables

| | (1) | (2) | (3) | (4) | (5) | (6) | (7) | (8) | (9) | (10) | (11) | (12) | (13) | (14) | (15) | (16) | (17) | (18) | (19) | (20) | (21) | (22) |
|---|---|---|---|---|---|---|---|---|---|---|---|---|---|---|---|---|---|---|---|---|---|---|
| (1) Number of Authors | 1 | | | | | | | | | | | | | | | | | | | | | |
| (2) Average Career Age | 0.14 | 1 | | | | | | | | | | | | | | | | | | | | |
| (3) Log(Cumulative Citations) | 0.25 | 0.61 | 1 | | | | | | | | | | | | | | | | | | | |
| (4) Number of Institutions | 0.39 | 0.14 | 0.20 | 1 | | | | | | | | | | | | | | | | | | |
| (5) Log(All Paper Counts) | 0.11 | 0.11 | 0.22 | 0.32 | 1 | | | | | | | | | | | | | | | | | |
| (6) Institution Dispersion (1-HHI) | 0.39 | 0.15 | 0.20 | 0.87 | 0.17 | 1 | | | | | | | | | | | | | | | | |
| (7) Number of Countries | 0.26 | 0.11 | 0.16 | 0.65 | 0.23 | 0.59 | 1 | | | | | | | | | | | | | | | |
| (8) Country Dispersion (1-HHI) | 0.19 | 0.11 | 0.14 | 0.53 | 0.09 | 0.60 | 0.93 | 1 | | | | | | | | | | | | | | |
| (9) Number of Level 0 Concepts | -0.03 | -0.09 | -0.04 | -0.02 | 0.04 | -0.03 | -0.02 | -0.03 | 1 | | | | | | | | | | | | | |
| (10) Number of Level 1 Concepts | -0.03 | -0.07 | -0.01 | 0.01 | 0.09 | 0.00 | 0.02 | 0.01 | 0.70 | 1 | | | | | | | | | | | | |
| (11) Number of Level 2 Concepts | -0.09 | -0.06 | -0.02 | -0.01 | 0.08 | -0.02 | 0.01 | 0.00 | 0.54 | 0.71 | 1 | | | | | | | | | | | |
| (12) One Year Citations | 0.17 | 0.10 | 0.20 | 0.13 | 0.11 | 0.10 | 0.13 | 0.11 | 0.00 | 0.02 | 0.02 | 1 | | | | | | | | | | |
| (13) Two Year Citations | 0.17 | 0.07 | 0.22 | 0.14 | 0.12 | 0.12 | 0.13 | 0.12 | 0.02 | 0.07 | 0.08 | 0.85 | 1 | | | | | | | | | |
| (14) Three Year Citations | 0.15 | 0.01 | 0.21 | 0.12 | 0.10 | 0.11 | 0.11 | 0.10 | 0.03 | 0.09 | 0.11 | 0.73 | 0.94 | 1 | | | | | | | | |
| (15) Paper Generation Score | 0.07 | 0.13 | 0.06 | 0.05 | 0.03 | 0.05 | 0.04 | 0.03 | -0.03 | -0.02 | -0.06 | -0.03 | -0.06 | -0.11 | 1 | | | | | | | |

| | (1) | (2) | (3) | (4) | (5) | (6) | (7) | (8) | (9) | (10) | (11) | (12) | (13) | (14) | (15) | (16) | (17) | (18) | (19) | (20) | (21) | (22) |
|---|---|---|---|---|---|---|---|---|---|---|---|---|---|---|---|---|---|---|---|---|---|---|
| (16) Publication Year | 0.10 | 0.40 | 0.13 | 0.07 | 0.05 | 0.04 | 0.06 | 0.05 | -0.16 | -0.18 | -0.23 | -0.06 | -0.20 | -0.36 | 0.29 | 1 | | | | | | |
| (17) Venue: Conference | -0.03 | -0.15 | -0.03 | -0.07 | -0.06 | -0.03 | -0.04 | -0.00 | 0.06 | 0.04 | 0.03 | -0.11 | -0.07 | -0.01 | -0.02 | -0.27 | 1 | | | | | |
| (18) Venue: Core Collection | 0.05 | 0.14 | 0.05 | 0.09 | 0.10 | 0.03 | 0.04 | -0.01 | -0.05 | -0.00 | -0.00 | 0.12 | 0.10 | 0.06 | 0.05 | 0.22 | -0.81 | 1 | | | | |
| (19) Venue: Impact Factor | 0.11 | 0.16 | 0.18 | 0.12 | 0.14 | 0.08 | 0.11 | 0.08 | -0.02 | 0.01 | 0.03 | 0.30 | 0.30 | 0.25 | 0.05 | 0.15 | -0.33 | 0.33 | 1 | | | |
| (20) Any Industrial Affiliation | 0.20 | 0.09 | 0.10 | 0.17 | 0.08 | 0.16 | 0.14 | 0.12 | -0.00 | 0.01 | 0.03 | 0.13 | 0.15 | 0.16 | -0.02 | -0.08 | 0.07 | -0.08 | 0.04 | 1 | | |
| (21) Previous BERT Publications | 0.02 | 0.15 | 0.26 | 0.05 | 0.09 | 0.08 | 0.08 | 0.10 | 0.03 | 0.02 | 0.04 | 0.16 | 0.13 | 0.09 | 0.06 | 0.11 | 0.05 | -0.03 | 0.10 | 0.06 | 1 | |
| (22) Dummy: Only Mention BERT | -0.05 | -0.09 | -0.06 | -0.05 | -0.04 | -0.05 | -0.04 | -0.03 | 0.02 | 0.01 | 0.03 | 0.02 | 0.02 | 0.06 | -0.83 | -0.25 | 0.04 | -0.06 | -0.05 | 0.01 | -0.06 | 1 |

4 Results

## 4.1 Larger collaborative teams and more experienced authors are involved in the development of later generations of models

We first tested Hypothesis 1A (Newer BERT-family models involve more collaborators), focusing on the number and experience of authors within the research team. The models are reported in Table 3. For each of our three variables, we estimated three models. In addition to the main independent variable (generation score), the first model controls only for the broad domain of the paper; the second includes all major control variables except objective publication time; and the third adds objective publication time, which represents one of the most important confounders for generation score. Our findings suggest that later-generation BERT models are generally developed by larger teams that include more experienced researchers. According to the estimates from our most rigorous models (Models 3, 6, and 9 in Table 3), an increase in the generation score from its minimum (0) to maximum (14.88) is associated with 0.31 additional authors ($14.88 \times 0.021 = 0.31$), each having 1.22 more years of research experience ($14.88 \times 0.082 = 1.22$). These conclusions are supported by several sets of controls, which demonstrate the robustness of our conclusion.

Our findings align with the well-documented trend of increasing collaboration sizes across research fields (Cronin, 2021; Larivière et al., 2015; Wuchty et al., 2007). By including publication year fixed effects in our most rigorous models (Models 3, 6, and 9), we demonstrate that these effects persist even after accounting for broader temporal changes in academic culture. To further compare the magnitude of the generation effect and the period effect (i.e., publication year), we include both variables in the model and compare their standardized regression coefficients. These results are presented in Table 8. Note that Table 8 reports standardized coefficients, whereas other tables present unstandardized ones, so the numbers can be different. Standardized coefficients allow for direct comparison of the relative importance of variables, as they indicate how many standard deviations the dependent variable will change in response to a one standard deviation change in the independent variable.

Table 8 shows that both the significance and direction of our coefficients remain consistent after accounting for publication year. While the generation score has a slightly greater effect size than publication year in predicting the number of authors, publication year has a larger effect size when predicting author experience.

One possible concern is that the generation score might capture time effects not absorbed by publication year (e.g., differences between models published in January versus October of the same year). To address this, we conducted a series of robustness tests, reported in Appendix Tables 18–20, using different time windows to control for objective publication time. The effect sizes of generation score remain consistent across these specifications, indicating that the effects of model generation are largely independent of objective time effects.

As for the control variables, larger and more experienced teams are also associated with higher article impact, as indicated by the positive coefficients for *Quality: One-Year Citations*, and with publication in higher-impact venues, as shown by the positive coefficients for *Venue: Impact Factor*. In addition, our results show that our key findings persist even when accounting for industrial participation (Tables 18-20 in the Appendix).

Next, we tested Hypothesis 1B (Newer BERT-family models involve more institutions), as reported in Table 4. While the raw models (Models 1, 4, and 7) show positive and significant coefficients, these effects become insignificant after adding additional controls (Models 3, 6, and 9). The only dependent variable for which the generation score shows marginal significance in the most rigorous models is institutional dispersion ($\beta = 0.008+$ in Model 9). This suggests that an increase in the generation score from its minimum (0) to maximum (14.88) is associated with a 0.119-point increase in the institutional dispersion index ($14.88 \times 0.008 = 0.119$). Regarding controls, our models in Table 4 also suggest that higher-quality papers published in high-impact journals or conferences tend to have authors who are more evenly distributed across multiple institutions.

Finally, we examine how international collaborations have evolved across different generations of BERT algorithms. The international collaboration around AI research is an interesting topic that is being paid increasing attention (AlShebli et al., 2024; Hu et al., 2020). Using the same model specifications as in Tables 3 and 4, we examined whether later-generation models are more likely to include authors from multiple countries and whether the authors are more evenly distributed across these countries. Interestingly, the results in Table 5 do not support the claim that international collaboration has increased in the development of AI models. Although all coefficients are positive, none are statistically significant. While Table 8 shows that BERT papers published in more recent years tend to involve more international collaboration, this effect does not hold for generation scores.

To briefly summarize: Our models provide clear evidence supporting Hypothesis 1A, offer only partial and marginal support for Hypothesis 1B. Furthermore, we found no evidence of an increase in international collaborations.

| | DV: Number of Authors | | | DV: Average Career Age | | | DV: log (Cumulative Citations) | | |
|---|---|---|---|---|---|---|---|---|---|
| | Model 1 | Model 2 | Model 3 | Model 4 | Model 5 | Model 6 | Model 7 | Model 8 | Model 9 |
| Paper Generation | 0.015*** | 0.026*** | 0.021** | 0.129*** | 0.185*** | 0.082** | 0.052*** | 0.038 | 0.022 |
| | (0.004) | (0.006) | (0.006) | (0.017) | (0.029) | (0.028) | (0.014) | (0.024) | (0.024) |
| Quality: One Year Citations | | 0.014*** | 0.014*** | | 0.017+ | 0.033*** | | 0.057*** | 0.063*** |
| | | (0.002) | (0.002) | | (0.009) | (0.009) | | (0.007) | (0.008) |
| Number of Level 2 Concepts | | -0.019*** | -0.019*** | | -0.025 | 0.035* | | -0.026+ | -0.013 |
| | | (0.004) | (0.004) | | (0.017) | (0.017) | | (0.014) | (0.014) |
| Venue: Conference | | 0.066* | 0.132*** | | -0.499*** | 0.101 | | 0.201+ | 0.400*** |
| | | (0.032) | (0.034) | | (0.142) | (0.143) | | (0.115) | (0.121) |
| Venue: Core Collection | | 0.091* | 0.130*** | | 0.186 | 0.264+ | | 0.226+ | 0.281* |
| | | (0.035) | (0.037) | | (0.158) | (0.155) | | (0.128) | (0.132) |
| Venue: Impact Factor | | 0.006* | 0.006* | | 0.052*** | 0.042*** | | 0.055*** | 0.062*** |
| | | (0.002) | (0.003) | | (0.011) | (0.011) | | (0.009) | (0.010) |
| Any Industrial Affiliation | | 0.235*** | 0.238*** | | 0.564*** | 0.686*** | | 0.365*** | 0.386*** |
| | | (0.020) | (0.020) | | (0.098) | (0.093) | | (0.079) | (0.079) |
| Previous BERT Publications | | -0.003 | -0.007 | | 0.388*** | 0.307*** | | 0.511*** | 0.527*** |
| | | (0.009) | (0.010) | | (0.043) | (0.042) | | (0.035) | (0.036) |
| Field Fixed Effects | YES | YES | YES | YES | YES | YES | YES | YES | YES |
| Publication Time Fixed Effects | | | YES | | | YES | | | YES |
| Constant | 1.371*** | 1.260*** | 1.135*** | 6.322*** | 6.185*** | 4.280*** | 6.190*** | 5.767*** | 5.212*** |
| | (0.029) | (0.051) | (0.063) | (0.137) | (0.229) | (0.267) | (0.114) | (0.186) | (0.227) |
| Adjusted R-squared/Pseudo R2 | 0.007 | 0.022 | 0.026 | 0.041 | 0.098 | 0.199 | 0.025 | 0.125 | 0.139 |
| N | 4130 | 4130 | 4130 | 4121 | 4121 | 4121 | 4121 | 4121 | 4121 |

*Note: Standard errors in parentheses. $^{+}p < 0.1$, $^{*}p < 0.05$, $^{**}p < 0.01$, $^{***}p < 0.001$. All tests are two-tailed.*

Table 4. Regression Estimates of Generation Effects on Research Inputs: Institutions

| | DV: Number of Institutions | | | DV: log (All Paper Counts) | | | DV: Institution Dispersion | | |
|---|---|---|---|---|---|---|---|---|---|
| | Model 1 | Model 2 | Model 3 | Model 4 | Model 5 | Model 6 | Model 7 | Model 8 | Model 9 |
| Paper Generation | 0.012* | 0.012 | 0.011 | 0.034** | 0.015 | 0.006 | 0.006** | 0.007+ | 0.008+ |
| | (0.006) | (0.010) | (0.010) | (0.012) | (0.020) | (0.020) | (0.002) | (0.004) | (0.004) |
| Quality: One Year Citations | | 0.010*** | 0.010*** | | 0.024*** | 0.026*** | | 0.004*** | 0.004** |
| | | (0.003) | (0.003) | | (0.006) | (0.006) | | (0.001) | (0.001) |
| Number of Level 2 Concepts | | -0.001 | -0.002 | | -0.016 | -0.008 | | -0.000 | -0.001 |
| | | (0.006) | (0.006) | | (0.012) | (0.012) | | (0.002) | (0.002) |
| Venue: Conference | | 0.047 | 0.086+ | | 0.043 | 0.170 | | 0.017 | 0.028 |
| | | (0.049) | (0.052) | | (0.101) | (0.107) | | (0.020) | (0.021) |
| Venue: Core Collection | | 0.141** | 0.163** | | -0.001 | 0.023 | | 0.032 | 0.039+ |
| | | (0.054) | (0.056) | | (0.110) | (0.114) | | (0.022) | (0.023) |
| Venue: Impact Factor | | 0.009* | 0.010* | | 0.029*** | 0.030*** | | 0.003+ | 0.003* |
| | | (0.004) | (0.004) | | (0.008) | (0.008) | | (0.001) | (0.002) |
| Any Industrial Affiliation | | 0.274*** | 0.274*** | | -0.931*** | -0.926*** | | 0.116*** | 0.115*** |
| | | (0.031) | (0.031) | | (0.064) | (0.064) | | (0.013) | (0.013) |
| Previous BERT Publications | | 0.022 | 0.020 | | 0.175*** | 0.174*** | | 0.020*** | 0.020*** |
| | | (0.014) | (0.015) | | (0.029) | (0.030) | | (0.006) | (0.006) |
| Field Fixed Effects | YES | YES | YES | YES | YES | YES | YES | YES | YES |
| Publication Time Fixed Effects | | | YES | | | YES | | | YES |
| Constant | 0.534*** | 0.374*** | 0.275** | 10.052*** | 10.175*** | 9.719*** | 0.252*** | 0.187*** | 0.169*** |
| | (0.046) | (0.078) | (0.098) | (0.097) | (0.162) | (0.197) | (0.019) | (0.032) | (0.039) |
| Adjusted R-squared/Pseudo R2 | 0.004 | 0.015 | 0.016 | 0.013 | 0.085 | 0.095 | 0.013 | 0.049 | 0.049 |
| N | 4130 | 4130 | 4130 | 3550 | 3550 | 3550 | 3550 | 3550 | 3550 |

Note: Standard errors in parentheses. $^+ p < 0.1$, $^* p < 0.05$, $^{**} p < 0.01$, $^{***} p < 0.001$. All tests are two-tailed.

Table 5. Regression Estimates of Generation Effects on Research Inputs: International Collaborations

| | DV: Number of Countries Involved | | | DV: Country Dispersion | | |
|---|---|---|---|---|---|---|
| | Model 1 | Model 2 | Model 3 | Model 4 | Model 5 | Model 6 |
| Paper Generation | 0.007 | 0.006 | 0.004 | 0.002 | 0.002 | 0.002 |
| | (0.006) | (0.012) | (0.012) | (0.002) | (0.003) | (0.003) |
| Quality: One Year Citations | | 0.009* | 0.010** | | 0.004*** | 0.004*** |
| | | (0.003) | (0.004) | | (0.001) | (0.001) |
| Number of Level 2 Concepts | | 0.003 | 0.004 | | 0.001 | 0.001 |
| | | (0.007) | (0.007) | | (0.002) | (0.002) |
| Venue: Conference | | 0.010 | 0.038 | | -0.000 | 0.009 |
| | | (0.056) | (0.060) | | (0.015) | (0.016) |
| Venue: Core Collection | | 0.030 | 0.042 | | -0.011 | -0.008 |
| | | (0.062) | (0.065) | | (0.016) | (0.017) |
| Venue: Impact Factor | | 0.007+ | 0.007 | | 0.003* | 0.003* |
| | | (0.004) | (0.005) | | (0.001) | (0.001) |
| Any Industrial Affiliation | | 0.151*** | 0.154*** | | 0.057*** | 0.058*** |
| | | (0.037) | (0.037) | | (0.009) | (0.009) |
| Previous BERT Publications | | 0.029+ | 0.024 | | 0.019*** | 0.017*** |
| | | (0.016) | (0.017) | | (0.004) | (0.004) |
| Field Fixed Effects | YES | YES | YES | YES | YES | YES |
| Publication Time Fixed Effects | | | YES | | | YES |
| Constant | 0.239*** | 0.143 | 0.052 | 0.095*** | 0.060* | 0.035 |
| | (0.053) | (0.091) | (0.114) | (0.014) | (0.024) | (0.029) |
| Adjusted R-squared/Pseudo R2 | 0.001 | 0.006 | 0.006 | 0.013 | 0.042 | 0.041 |
| N | 4130 | 4130 | 4130 | 3550 | 3550 | 3550 |

*Note: Standard errors in parentheses.* $^+ p < 0.1,$ *$^* p < 0.05,$* *$^{**} p < 0.01,$* *$^{***} p < 0.001.$ *All tests are two-tailed.*

**4.2 Later generations of BERT models are more specialized.**

We then discuss how model generation is linked to the breadth of concepts in BERT-related papers. We regress the number of concepts on the generation score, with all controls in place. The results, as reported in Table 6, show that later-generation models involve fewer concepts on average. These conclusions hold for most rigorous models ($\beta$ = -0.008* in Model 3 for level 0 concepts, and $\beta$ = -0.058* in Model 9 for level 2 concepts). Therefore, compared with early BERT models, later-generation models tend to be more specialized, focusing on narrower questions within local communities. Hypothesis 2 is supported, i.e., publications on newer BERT-family models have narrower conceptual scopes.

The standardized regression coefficients in Table 8 show that the trend toward specialization applies not only to later-generation studies but also to those published in more recent years. In fact, publication year has a larger effect size than generation score for this outcome.

Knowledge breadth can be seen as a signal of breaking disciplinary boundaries or integrating insights from diverse fields (Rafols & Meyer, 2010). From this perspective, our findings are consistent with theoretical arguments that seminal works often challenge established intellectual paradigms, while later studies tend to align with the norms and practices of existing scientific disciplines (Kuhn, 1996). This pattern also reflects the growing specialization of science, a trend widely discussed in classic studies of scientific development (Ben-David & Collins, 1966; Price, 1969; Wray, 2005). It also echoes recent empirical evidence that scientific research has become less disruptive over time (Park et al., 2023).

Table 6. Regression Estimates of Generation Effects on Knowledge Breadth

| | DV: Level 0 Concepts | | | DV: Level 1 Concepts | | | DV: Level 2 Concepts | | |
|---|---|---|---|---|---|---|---|---|---|
| | Model 1 | Model 2 | Model 3 | Model 4 | Model 5 | Model 6 | Model 7 | Model 8 | Model 9 |
| Paper Generation | -0.003 | -0.011** | -0.008* | -0.008 | -0.023 | -0.005 | -0.055*** | -0.107*** | -0.058* |
| | (0.002) | (0.004) | (0.004) | (0.010) | (0.019) | (0.019) | (0.015) | (0.027) | (0.026) |
| Quality: One Year Citations | | 0.002 | 0.001 | | 0.012* | 0.009 | | 0.018* | 0.009 |
| | | (0.001) | (0.001) | | (0.006) | (0.006) | | (0.008) | (0.008) |
| Number of Authors | | 0.001 | 0.002 | | -0.010 | -0.009 | | -0.097*** | -0.091*** |
| | | (0.002) | (0.002) | | (0.012) | (0.012) | | (0.017) | (0.017) |
| Number of Institutions | | 0.004 | 0.004 | | 0.045* | 0.045* | | 0.037 | 0.034 |
| | | (0.004) | (0.004) | | (0.021) | (0.021) | | (0.030) | (0.029) |
| Venue: Conference | | 0.058** | 0.045* | | 0.293** | 0.326*** | | 0.281* | 0.274* |
| | | (0.018) | (0.019) | | (0.090) | (0.095) | | (0.128) | (0.133) |
| Venue: Core Collection | | 0.012 | 0.011 | | 0.333*** | 0.476*** | | 0.333* | 0.566*** |
| | | (0.020) | (0.020) | | (0.101) | (0.103) | | (0.143) | (0.144) |
| Venue: Impact Factor | | -0.001 | -0.001 | | 0.006 | 0.001 | | 0.015 | 0.013 |
| | | (0.001) | (0.001) | | (0.007) | (0.008) | | (0.010) | (0.011) |
| Any Industrial Affiliation | | 0.016 | 0.011 | | 0.082 | 0.026 | | 0.237** | 0.123 |
| | | (0.013) | (0.013) | | (0.065) | (0.064) | | (0.092) | (0.090) |
| Previous BERT Publications | | 0.012* | 0.012* | | 0.006 | -0.018 | | 0.050 | 0.035 |
| | | (0.005) | (0.006) | | (0.028) | (0.028) | | (0.039) | (0.039) |
| Field Fixed Effects | YES | YES | YES | YES | YES | YES | YES | YES | YES |
| Publication Time Fixed Effects | | | YES | | | YES | | | YES |
| Constant | 2.783*** | 2.756*** | 2.763*** | 5.178*** | 4.868*** | 4.683*** | 5.193*** | 5.355*** | 5.381*** |
| | (0.017) | (0.028) | (0.034) | (0.086) | (0.141) | (0.172) | (0.122) | (0.200) | (0.241) |
| Adjusted R-squared/Pseudo R2 | 0.951 | 0.952 | 0.952 | 0.483 | 0.485 | 0.499 | 0.335 | 0.343 | 0.381 |
| N | 4130 | 4130 | 4130 | 4130 | 4130 | 4130 | 4130 | 4130 | 4130 |

*Note: Standard errors in parentheses. $^+ p < 0.1$, $^* p < 0.05$, $^{**} p < 0.01$, $^{***} p < 0.001$. All tests are two-tailed.*

### 4.3 Lower citation impact, driven by longer-term citation impact, of later generations of models

Finally, we examine the relationship between generation score and research impact. We examine the citation impact received by BERT studies based on negative binomial regressions, considering different time windows for citations (i.e., one to three years). The results are reported in Table 7.

The most rigorous models in Table 7 show that newer models tend to have lower research impact on average (β = -0.034* in Model 3 for one-year citations, β = -0.075*** in Model 6 for two-year citations, and β = -0.102*** in Model 9 for three-year citations). The coefficients suggest that the citation gap between older and newer models widens over time. An increase in the generation score from the empirical minimum (0) to the maximum (14.88) reduces one-year citations by only 0.51 (–0.034 × 14.88 = –0.51), but leads to a decline of 1.52 in three-year citations (–0.102 × 14.88 = –1.52). When comparing the effect sizes of publication year and generation score in Table 8, we find that generation score has a greater effect. This suggests that the citation advantage of earlier-generation models stems more from their pioneering position in the knowledge trajectory, than from simply being published earlier. Hence, the results validate Hypothesis 3 (Newer BERT-family models receive fewer citations).

This pattern is also supported by the descriptive results in Figure 2. In this figure, we divide all samples into ten equal-sized quantile groups based on generation scores and calculate their average citations, allowing us to visualize citation differences without additional controls. While BERT papers show relatively similar one-year citation impact, the gap becomes more pronounced in three-year citations: the earliest models receive approximately 50% more citations than those in the latest generation quantile. This gradient is also evident in the regression results, as shown in Table 16 in the Appendix B.

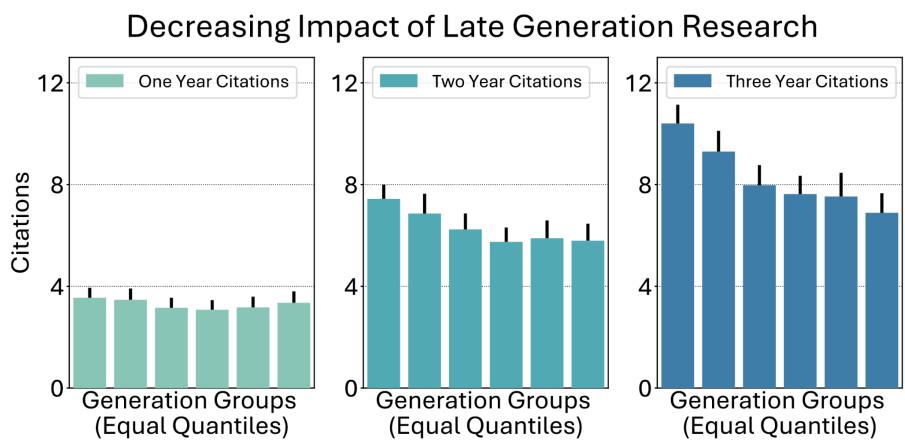

Figure 2. Description of Citation Gaps Between Early and Late-Generation BERT Models

Table 7. Regression Estimates of Generation Effects on Research Impact: Citations

| | DV: One Year Forward Citations (Paper in and before 2023) | | | DV: Two Year Forward Citations (Paper in and before 2022) | | | DV: Three Year Forward Citations (Paper in and before 2021) | | |
|---|---|---|---|---|---|---|---|---|---|
| | Model 1 | Model 2 | Model 3 | Model 4 | Model 5 | Model 6 | Model 7 | Model 8 | Model 9 |
| Paper Generation | -0.009 | -0.039* | -0.034* | -0.015+ | -0.093*** | -0.075*** | -0.021* | -0.117*** | -0.102*** |
| | (0.009) | (0.016) | (0.016) | (0.008) | (0.015) | (0.015) | (0.009) | (0.019) | (0.018) |
| Number of Authors | | 0.072*** | 0.072*** | | 0.055*** | 0.054*** | | 0.051*** | 0.046*** |
| | | (0.010) | (0.010) | | (0.009) | (0.009) | | (0.010) | (0.010) |
| Number of Institutions | | 0.013 | 0.014 | | 0.037* | 0.042** | | 0.022 | 0.020 |
| | | (0.018) | (0.017) | | (0.015) | (0.015) | | (0.017) | (0.017) |
| Number of Level 2 Concepts | | 0.004 | 0.005 | | 0.020* | 0.018* | | 0.017* | 0.009 |
| | | (0.009) | (0.009) | | (0.008) | (0.008) | | (0.009) | (0.009) |
| Venue: Conference | | 0.065 | 0.245** | | 0.160* | 0.357*** | | 0.157+ | 0.354*** |
| | | (0.080) | (0.084) | | (0.072) | (0.080) | | (0.082) | (0.094) |
| Venue: Core Collection | | 0.203* | 0.409*** | | 0.239** | 0.450*** | | 0.225* | 0.451*** |
| | | (0.087) | (0.090) | | (0.080) | (0.086) | | (0.091) | (0.101) |
| Venue: Impact Factor | | 0.072*** | 0.071*** | | 0.054*** | 0.060*** | | 0.038*** | 0.044*** |
| | | (0.006) | (0.006) | | (0.005) | (0.005) | | (0.006) | (0.006) |
| Any Industrial Affiliation | | 0.228*** | 0.183*** | | 0.206*** | 0.167*** | | 0.184*** | 0.167*** |
| | | (0.050) | (0.049) | | (0.042) | (0.042) | | (0.045) | (0.044) |
| Previous BERT Publications | | 0.133*** | 0.160*** | | 0.085*** | 0.122*** | | 0.090*** | 0.155*** |
| | | (0.021) | (0.021) | | (0.020) | (0.020) | | (0.025) | (0.026) |
| Field Fixed Effects | YES | YES | YES | YES | YES | YES | YES | YES | YES |
| Publication Time Fixed Effects | | | YES | | | YES | | | YES |
| Constant | 1.241*** | 0.718*** | 1.266*** | 2.057*** | 1.632*** | 1.604*** | 1.805*** | 1.561*** | 1.517*** |
| | (0.073) | (0.131) | (0.145) | (0.059) | (0.115) | (0.130) | (0.390) | (0.402) | (0.403) |
| Adjusted R-squared/Pseudo R2 | 0.004 | 0.026 | 0.044 | 0.004 | 0.023 | 0.030 | 0.004 | 0.017 | 0.024 |

*Note: Standard errors in parentheses. $^+ p < 0.1$, $^* p < 0.05$, $^{**} p < 0.01$, $^{***} p < 0.001$. All tests are two-tailed.*

Table 8. Standardized Regression Coefficients of Publication Year and Generation

| IV\DV | Num. of Authors | Ave. Career Age | log(Cumulative Citations) | Number of Institutions | log(Paper Count) | Institution Dispersion | Num. of Countries | Country Dispersion |
|---|---|---|---|---|---|---|---|---|
| Publication Year | 0.021*** | 0.372*** | 0.096*** | 0.025* | 0.100*** | 0.007 | 0.046 | 0.038* |
| | (4.87) | (23.26) | (5.77) | (2.10) | (5.50) | (0.40) | (1.73) | (2.03) |
| Paper Generation | 0.022** | 0.075** | 0.018 | 0.017 | -0.001 | 0.053 | 0.010 | 0.015 |
| | (3.27) | (2.92) | (0.69) | (0.92) | (-0.04) | (1.74) | (0.23) | (0.49) |

| IV\DV | Level 0 Concepts | Level 1 Concepts | Level 2 Concepts | One Year Citations | Two Year Citations | Three Year Citations |
|---|---|---|---|---|---|---|
| Publication Year | -0.017*** | -0.079*** | -0.156*** | -0.032*** | -0.026*** | -0.019*** |
| | (-4.29) | (-6.26) | (-11.03) | (-6.36) | (-8.72) | (-8.05) |
| Paper Generation | -0.013* | -0.004 | -0.051* | -0.009 | -0.026*** | -0.026*** |
| | (-2.13) | (-0.20) | (-2.22) | (-1.07) | (-5.00) | (-5.44) |

Our finding aligns with the observation that breakthrough studies tend to have greater scientific impact than incremental ones (Schilling & Green, 2011), particularly in the long term (Cao & Evans, 2024). More importantly, the differing citation patterns between newer and older models appear to reflect the two types of citation functions described by Leydesdorff et al. (2016): short-term citations indicate the relevance to the current research front, while long-term citations signify the codification of knowledge into conceptual symbols. This connection offers new insights into the recognition patterns of AI research. Early models, such as RoBERTa, are cited not only for their technical contributions but also for their symbolic significance in the process of knowledge production.

**4.4 Robustness Tests**

When a paper mentions only "BERT" in its abstract, it may be only marginally related to BERT algorithms (e.g., using BERT as a benchmark). Such studies may differ in nature from those that engage more directly with BERT development. To address this potential issue and reduce ambiguity, we conducted a robustness check by excluding these papers from the analysis, resulting in a subsample of 2,743 records. We then estimated the regression models using the same specifications as in the main analysis.

The results, presented in Appendix B, are consistent with the main findings: all hypotheses are supported in the regressions conducted on this subsample. Therefore, we argue that our conclusions are robust and not influenced by sample selection.

We also conducted several additional robustness tests by slightly adjusting the measurement construction and time window to examine whether our model results remain consistent under different conditions. Details of these tests and their results are presented in Appendix B. Overall, our findings are highly robust, with consistent empirical patterns observed across different variable constructions and model specifications.

5 Discussion

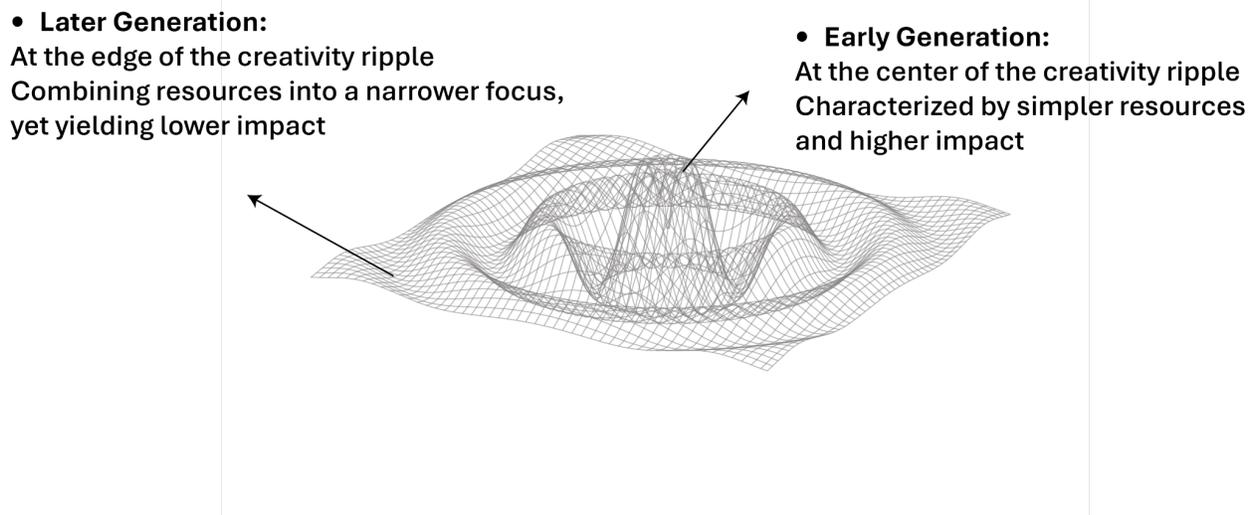

• **Later Generation:**
At the edge of the creativity ripple
Combining resources into a narrower focus,
yet yielding lower impact

• **Early Generation:**
At the center of the creativity ripple
Characterized by simpler resources
and higher impact

Figure 3. Conceptual illustration of Diminishing Returns to Research Input in AI Development

This study examines the production and impact of AI technologies through the case of BERT-family models, focusing on the resources used in their development—measured by the number and nature of authors and institutions involved—and the subsequent citations received within three years of publication. While newer BERT models require more resources for their development, they tend to receive fewer long-term citations than their predecessors. This is contrasting to the old wisdom that incremental improvements are normally simpler and hence only require smaller collaboration teams (Beaver & Rosen, 1978). And it also supplies a valuable data point to the finding in science of science research that there is a general negative correlation between the team size and innovation of research (Tang et al., 2024; Wu et al., 2019). In addition, our results show that the initial models in the family also received more citations than their descendants. This finding offers validation to the "first-mover advantage" proposed by Newman (2009) in the space of BERT-family models. Combining these findings together, we can draw the conclusion that early developments within a topic enjoy a higher return-on-investment. We illustrate this idea in Figure 3.

However, the above perspective is inherently biased against incremental research, which plays crucial roles in the ecology of knowledge production and dissemination (Wu et al., 2019). In our case, many newer BERT models focus on refining, applying, and extending the classic models focusing on narrower use cases, bridging methodological advances with broader scientific communities. And as Newman (2009) stated, being highly cited does not mean that first publications on a topic always have the best quality. We would argue that, if incremental research continues to receive less recognition, it may discourage essential and valid refinements and applications that contribute to the evolution and application of AI technologies.

In the meantime, the variation in citation counts among BERT models is largely driven by long-term citations rather than those received within the first year. This finding adds extra nuances to our story: all models are initially recognized as part of the research front (Leydesdorff et al., 2016), actively discussed and integrated into scholarly discourse; however, earlier models are more likely to be cited as symbolic representations of research topics, accruing a disproportionate share of citations simply due to their historical precedence (Small, 1978). This phenomenon reflects a Matthew Effect in the distribution of academic attention within a family of technologies: early models continue to accumulate citations, often unrelated to their specific contributions, while newer models struggle to gain similar recognition in the longer term despite often representing substantial refinements and improvements. At the same time, earlier models often have much higher failure rates (Franco & Filson, 2006; Markides & Sosa, 2013), which further suggests that the Matthew Effect we observe can also be a reward for the risk-taking efforts by these early-movers.

Our two findings underscore the challenges faced by incremental innovations in AI and highlight the need for more equitable research evaluation frameworks that recognize both foundational breakthroughs and incremental advancements in technological development. Current citation-based metrics, while useful, may undervalue the role of sustained innovation in AI, potentially discouraging researchers from engaging in necessary refinements and domain-specific applications. Policymakers, funding agencies, and academic institutions should explore complementary assessment mechanisms—such as impact case studies, qualitative peer reviews, and industry adoption evaluation—to provide a more holistic understanding of research contributions of these models.

6 Conclusion

This study provides a comprehensive analysis of the production and impact of BERT-family models, illuminating key dynamics in AI research. Our findings confirm that newer BERT models are developed by larger, more experienced, and institutionally diverse teams, reflecting the increasing complexity and collaborative nature of AI development (Cronin, 2001; Larivière et al., 2015). Additionally, these models exhibit greater specialization in topics and receive fewer citations, particularly over the long term, highlighting a "first-mover advantage" where early models like BERT accrue disproportionate recognition (Newman, 2009).

These insights underscore the need for more equitable evaluation frameworks in AI research. While foundational models drive transformative change, incremental advancements are essential for practical applications and domain-specific progress. Policymakers and academic institutions should consider complementary metrics, such as industry adoption or qualitative impact assessments, to better recognize the contributions of newer models. Furthermore, the resource-intensive nature of modern AI development calls for initiatives like open-access datasets and shared computational resources to promote inclusivity and reduce disparities in global research participation. Notably, all generations of BERT models receive similar citation counts in their first year. Although citation context is not the focus of this study, a plausible explanation for the gap between one-year and three-year citations is that earlier models garner more symbolic citations (Small, 1978), which may not directly reflect their scientific or practical values.

Despite its contributions, this study has limitations. First, it focuses solely on BERT-family models. While we believe these findings reflect broader patterns in AI research and development, future studies should compare different AI technologies to derive more generalizable, data-driven conclusions about their production and recognition in science. Second, the PWC and OpenAlex datasets have known limitations, such as incomplete institutions (Zhang et al., 2024) and inaccurate and missing citations (Culbert et al., 2025; Rodrigues et al., 2025) in OpenAlex, which may affect our analysis. Future research should leverage more comprehensive data sources and advanced methods for identifying AI models in publications (Gao & Wang, 2024) to build a more robust dataset to represent AI developments and their impacts on science. These steps will enhance our understanding of how AI development and recognition are embedded within the scientific system, a critical aspect of the science-technology relationship.

**Declaration**

The authors have no relevant financial or non-financial interests to disclose.

**REFERENCES**

Alcacer, J., & Gittelman, M. (2006). Patent citations as a measure of knowledge flows: The influence of examiner citations. *The review of economics and statistics*, *88*(4), 774-779.

AlShebli, B., Memon, S. A., Evans, J. A., & Rahwan, T. (2024). China and the US produce more impactful AI research when collaborating together. *Scientific Reports*, *14*(1), 28576.

Beaver, D., & Rosen, R. (1978). Studies in scientific collaboration: Part I. The professional origins of scientific co-authorship. *Scientometrics*, *1*(1), 65-84.

Beltagy, I., Lo, K., & Cohan, A. (2019). SciBERT: A pretrained language model for scientific text. *arXiv preprint arXiv:1903.10676*.

Ben-David, J., & Collins, R. (1966). Social factors in the origins of a new science: The case of psychology. *American sociological review*, 451-465.

Bornmann, L., & Daniel, H. D. (2008). What do citation counts measure? A review of studies on citing behavior. *Journal of documentation*, *64*(1), 45-80.

Brooks, H. (1994). The relationship between science and technology. *Research policy*, *23*(5), 477-486.

Cao, L., & Evans, J. (2024). Depth versus breadth in the hierarchical recombination of technology. *Academy of Management Proceedings*, 2024(1). https://doi.org/10.5465/amproc.2024.10275abstract

Cronin, B. (2001). Hyperauthorship: A postmodern perversion or evidence of a structural shift in scholarly communication practices?. *Journal of the American Society for Information Science and Technology*, 52(7), 558-569.

Culbert, J. H., Hobert, A., Jahn, N., Haupka, N., Schmidt, M., Donner, P., & Mayr, P. (2025). Reference coverage analysis of OpenAlex compared to Web of Science and Scopus. *Scientometrics*, *130*(4), 2475-2492.

Devlin, J., Chang, M. W., Lee, K., & Toutanova, K. (2019). Bert: Pre-training of deep bidirectional transformers for language understanding. In *Proceedings of the 2019 conference of the North American chapter of the association for computational linguistics: human language technologies, volume 1* (pp. 4171-4186).

Foster, J. G., Rzhetsky, A., & Evans, J. A. (2015). Tradition and innovation in scientists' research strategies. *American sociological review*, *80*(5), 875-908.

Franco, A. M., & Filson, D. (2006). Spin-outs: knowledge diffusion through employee mobility. *The Rand journal of economics*, *37*(4), 841-860.

Gao, J., & Wang, D. (2024). Quantifying the use and potential benefits of artificial intelligence in scientific research. *Nature human behaviour*, *8*(12), 2281-2292.

Gazni, A., Sugimoto, C. R., & Didegah, F. (2012). Mapping world scientific collaboration: Authors, institutions, and countries. *Journal of the American Society for Information Science and Technology*, *63*(2), 323-335.

Hu, H., Wang, D., & Deng, S. (2020). Global Collaboration in Artificial Intelligence: Bibliometrics and Network Analysis from 1985 to 2019. *J. Data Inf. Sci.*, 5(4), 86-115.

Kang, D., Danziger, R. S., Rehman, J., & Evans, J. A. (2024). Limited diffusion of scientific knowledge forecasts collapse. *Nature Human Behaviour*, 1–9.

Kardas, M., Czapla, P., Stenetorp, P., Ruder, S., Riedel, S., Taylor, R., & Stojnic, R. (2020). AxCell: Automatic Extraction of Results from Machine Learning Papers. In B. Webber, T. Cohn, Y. He, & Y. Liu (Eds.), *Proceedings of the 2020 Conference on Empirical Methods in Natural Language Processing (EMNLP)* (pp. 8580–8594). Association for Computational Linguistics. https://doi.org/10.18653/v1/2020.emnlp-main.692

Kuhn, T. S. (1996). *The Structure of Scientific Revolutions* (3rd edition). University of Chicago Press.

Lazer, D., Pentland, A., Adamic, L., Aral, S., Barabási, A. L., Brewer, D., ... & Van Alstyne, M. (2009). Computational social science. *Science*, 323(5915), 721-723.

Lan, Z., Chen, M., Goodman, S., Gimpel, K., Sharma, P., & Soricut, R. (2019). Albert: A lite bert for self-supervised learning of language representations. arXiv pr*eprint arXiv:1909.11942*.

Langham-Putrow, A., Bakker, C., & Riegelman, A. (2021). Is the open access citation advantage real? A systematic review of the citation of open access and subscription-based articles. *PloS one*, *16*(6), e0253129.

Larivière, V., Gingras, Y., Sugimoto, C. R., & Tsou, A. (2015). Team size matters: Collaboration and scientific impact since 1900. *Journal of the Association for Information Science and Technology*, *66*(7), 1323-1332.

Lee, J., Yoon, W., Kim, S., Kim, D., Kim, S., So, C. H., & Kang, J. (2020). BioBERT: a pre-trained biomedical language representation model for biomedical text mining. *Bioinformatics*, *36*(4), 1234-1240.

Leydesdorff, L., Bornmann, L., Comins, J. A., & Milojević, S. (2016). Citations: Indicators of quality? The impact fallacy. *Frontiers in Research metrics and Analytics*, *1*, 1.

Liu, Y., Ott, M., Goyal, N., Du, J., Joshi, M., Chen, D., ... & Stoyanov, V. (2019). Roberta: A robustly optimized bert pretraining approach. *arXiv preprint arXiv:1907.11692*.

Markides, C., & Sosa, L. (2013). Pioneering and first mover advantages: the importance of business models. *Long range planning*, *46*(4-5), 325-334.

Moutsinas, G., Shuaib, C., Guo, W., & Jarvis, S. (2021). Graph hierarchy: a novel framework to analyse hierarchical structures in complex networks. *Scientific Reports*, *11*(1), 13943.

Newman, M. E. (2001). The structure of scientific collaboration networks. *Proceedings of the national academy of sciences*, *98*(2), 404-409.

Newman, M. E. (2009). The first-mover advantage in scientific publication. *Europhysics Letters*, *86*(6), 68001.

Noble, D. (2002). The rise of computational biology. *Nature Reviews Molecular Cell Biology*, *3*(6), 459-463.

Palmer, R., & Brookes, R. (2002). Incremental innovation: A case study analysis. *Journal of Database Marketing & Customer Strategy Management*, *10*, 71-83.

Park, M., Leahey, E., & Funk, R. J. (2023). Papers and patents are becoming less disruptive over time. *Nature*, 613(7942), 138–144.

Parolo, P. D. B., Pan, R. K., Ghosh, R., Huberman, B. A., Kaski, K., & Fortunato, S. (2015). Attention decay in science. *Journal of Informetrics*, *9*(4), 734-745.

Petersen, A. M., Fortunato, S., Pan, R. K., Kaski, K., Penner, O., Rungi, A., ... & Pammolli, F. (2014). Reputation and impact in academic careers. *Proceedings of the National Academy of Sciences*, *111*(43), 15316-15321.

Piwowar, H. A., & Vision, T. J. (2013). Data reuse and the open data citation advantage. *PeerJ*, *1*, e175.

Price, D. J. d. S. (1969). Little science, big science ... and beyond (2nd print). Columbia University Press.

Priem, J., Piwowar, H., & Orr, R. (2022). OpenAlex: A fully-open index of scholarly works, authors, venues, institutions, and concepts. *arXiv preprint arXiv:2205.01833*.

Rafols, I., & Meyer, M. (2010). Diversity and network coherence as indicators of interdisciplinarity: case studies in bionanoscience. *Scientometrics*, *82*(2), 263-287.

Rip, A. (1992). Science and technology as dancing partners. In *Technological development and science in the industrial age: New perspectives on the science-technology relationship* (pp. 231-270). Dordrecht: Springer Netherlands.

Rodrigues, D., Lopes, A., & Batista, F. (2025). Detecting incoherent citation data among three bibliometric platforms: OpenAlex, Scopus and Web of Science. *Journal of Information Science*, 01655515251330579.

Schilling, M. A., & Green, E. (2011). Recombinant search and breakthrough idea generation: An analysis of high impact papers in the social sciences. *Research Policy*, *40*(10), 1321-1331.

Sonnenwald, D. H. (2007). Scientific collaboration. *Annu. Rev. Inf. Sci. Technol.*, *41*(1), 643-681.

Small, H. G. (1978). Cited documents as concept symbols. *Social studies of science*, *8*(3), 327-340.

Stevens, R., Taylor, V., Nichols, J., Maccabe, A. B., Yelick, K., & Brown, D. (2020). *Ai for science: Report on the department of energy (doe) town halls on artificial intelligence (ai) for science* (No. ANL-20/17). Argonne National Lab.(ANL), Argonne, IL (United States).

Sugimoto, C. R., Sugimoto, T. J., Tsou, A., Milojević, S., & Larivière, V. (2016). Age stratification and cohort effects in scholarly communication: A study of social sciences. *Scientometrics*, *109*, 997-1016.

Tang, X., Li, X., & Yi, M. (2024). Investigating the causal effects of affiliation diversity on the disruption of papers in Artificial Intelligence. *Information Processing & Management*, *61*(5), 103806.

Xie, Y., Pan, Y., Xu, H., & Mei, Q. (2024). Bridging AI and Science: Implications from a Large-Scale Literature Analysis of AI4Science. *arXiv preprint arXiv:2412.09628*.

Xu, H., Liu, M., Bu, Y., Sun, S., Zhang, Y., Zhang, C., ... & Ding, Y. (2024). The impact of heterogeneous shared leadership in scientific teams. *Information Processing & Management*, *61*(1), 103542.

Wang, H., Fu, T., Du, Y., Gao, W., Huang, K., Liu, Z., ... & Zitnik, M. (2023). Scientific discovery in the age of artificial intelligence. *Nature*, *620*(7972), 47-60.

Wang, X., Liu, C., Mao, W., & Fang, Z. (2015). The open access advantage considering citation, article usage and social media attention. *Scientometrics*, *103*, 555-564.

Wray, K. B. (2005). Rethinking scientific specialization. *Social studies of science*, *35*(1), 151-164.

Wu, L., Wang, D., & Evans, J. A. (2019). Large teams develop and small teams disrupt science and technology. *Nature*, *566*(7744), 378-382.

Wuchty, S., Jones, B. F., & Uzzi, B. (2007). The increasing dominance of teams in production of knowledge. *Science*, *316*(5827), 1036-1039.

Zeng, A., Fan, Y., Di, Z., Wang, Y., & Havlin, S. (2021). Fresh teams are associated with original and multidisciplinary research. *Nature human behaviour*, *5*(10), 1314-1322.

Zhang, L., Cao, Z., Shang, Y., Sivertsen, G., & Huang, Y. (2024). Missing institutions in
OpenAlex: Possible reasons, implications, and solutions. *Scientometrics*, *129*(10), 5869-
5891.

# Appendices for *How BERT Proliferated*

## Appendix A. Method Details for Identifying True BERT Concepts

We employ a deep learning model to distinguish BERT-related terms (e.g., RoBERTa) from irrelevant nouns (e.g., Hilbert Space). To achieve this goal, we first fine-tune a bert-base-uncased model on all Papers with Code abstracts, enabling the algorithm to better capture information and identify logical relationships as expressed in computer science publications. Using this fine-tuned model, we generate contextual word embeddings for each BERT-related concept in every abstract. Contextual embeddings produce distinct vectors for the same word when it appears in different contexts. A concept is classified as a true BERT reference if its cosine similarity to the average embedding of "BERT" exceeds a threshold of 0.3, selected based on human reading of the corpus.

To assess the performance of this simple classifier, we validate it using a human-labeled dataset. Specifically, we manually annotate 500 sentences containing BERT-related phrases, extracted from PwC abstracts. Several examples are presented in Table 9.

Table 9. Examples of Human-Labelled BERT Terms

| Sentence | Label |
|---|---|
| • In contrast , sapBERT outperformed xlinker in the remaining  datasets. | True |
| • In vfas , vectors can represent individual data points as well as elements of a function space, a reproducing kernel hilBERT space. | False |
| • We performed experiments with the pubmedBERT pretrained model, which was fine tuned on a specific corpus for the task. | True |

We apply our classification approach to distinguish true BERT terms from other similarly spelled terms. Compared to the ground truth based on human annotations, our method achieves an accuracy of 0.97 and an F1 score of 0.95, demonstrating its effectiveness in identifying BERT-related concepts. Additional performance metrics are reported in Table 10.

Table 10. Classification Report for BERT Classifier

|  | Precision | Recall | F1-Score | N |
|---|---|---|---|---|
| 0 (Negative Sample) | 0.97 | 0.87 | 0.92 | 83 |
| 1 (Positive Sample) | 0.97 | 1.00 | 0.98 | 417 |
| Accuracy | | | 0.97 | 500 |
| Macro Avg | 0.97 | 0.93 | 0.95 | 500 |
| Weighted Avg | 0.97 | 0.97 | 0.97 | 500 |

**Appendix B. Robustness Tests**

To assess the robustness of our findings, we perform several additional robustness tests.

First, we exclude papers that mention only "BERT" in the abstract without referencing any other BERT-related terms. Among the original 4,208 records, 2,711 meet this inclusion criterion. The results, reported in Tables 11-Table 15, remain consistent with those from our baseline models.

Second, we replace the continuous generation scores with ordinal variables to examine their effects on paper impact. The results, shown in Table 16, reveal a gradual decline in impact as generation scores increase.

Third, we adopt an alternative approach to calculate generation scores. While we continue to use the graph hierarchy algorithm described in the main text, we estimate each paper's generation score directly from its citation network. The results, reported in Table 17, exhibit similar patterns in the research impact models, further reinforcing our conclusions. As expected, the effects on citations are more pronounced, as the measure is derived from citation data. Although the generation effects in the research input models are largely insignificant, their directions remain consistent.

Fourth, we re-estimated our models using different time windows for objective time as fixed effects. These robustness tests aim to demonstrate that the generation effect represents an independent influence that persists regardless of how objective time is measured. The results, reported in Tables 18–20, remain consistent with those presented in the main text.

Fifth, we conducted a robustness test on the citation models by removing all self-citations, ensuring that the higher citation counts of earlier models are not driven by the same research groups citing their own work. The results, reported in Table 21, display patterns consistent with those in Table 7, confirming the robustness of our findings.

Sixth, we conducted a robustness test using different thresholds for concept inclusion, setting the selection thresholds at 0.1, 0.2, and 0.5, respectively. We constructed knowledge breadth based on these different criteria and re-estimate the models. The results, reported in Table 22, are overall stable and consistent with those presented in the main text.

Finally, we conducted an additional analysis using industrial participation as the dependent variable to examine how this factor evolves with model development. The results are reported in Table 23.

Table 11. Robustness Test 1: Regression Estimates of Generation Effects on Author Characteristics, Based on a Selective Sample

| | DV: Number of Authors | | | DV: Average Career Age | | | DV: log (Cumulative Citations) | | |
|---|---|---|---|---|---|---|---|---|---|
| | Model 1 | Model 2 | Model 3 | Model 4 | Model 5 | Model 6 | Model 7 | Model 8 | Model 9 |
| Paper Generation | 0.026*** | 0.027*** | 0.022*** | 0.186*** | 0.189*** | 0.082** | 0.038 | 0.042+ | 0.021 |
| | (0.006) | (0.006) | (0.007) | (0.030) | (0.030) | (0.029) | (0.025) | (0.024) | (0.024) |
| Quality: One Year Citations | | 0.017*** | 0.017*** | | 0.035** | 0.042*** | | 0.063*** | 0.068*** |
| | | (0.002) | (0.002) | | (0.012) | (0.011) | | (0.009) | (0.009) |
| Number of Level 2 Concepts | | -0.023*** | -0.023*** | | -0.019 | 0.051* | | -0.015 | 0.003 |
| | | (0.005) | (0.005) | | (0.022) | (0.022) | | (0.017) | (0.018) |
| Venue: Conference | | 0.051 | 0.105* | | -0.542** | 0.019 | | -0.119 | 0.091 |
| | | (0.041) | (0.043) | | (0.188) | (0.186) | | (0.148) | (0.154) |
| Venue: Core Collection | | 0.078+ | 0.107* | | 0.116 | 0.148 | | -0.103 | -0.061 |
| | | (0.045) | (0.046) | | (0.206) | (0.199) | | (0.163) | (0.165) |
| Venue: Impact Factor | | 0.001 | 0.001 | | 0.039** | 0.036* | | 0.050*** | 0.057*** |
| | | (0.003) | (0.003) | | (0.014) | (0.014) | | (0.011) | (0.012) |
| Any Industrial Affiliation | | 0.212*** | 0.218*** | | 0.397** | 0.572*** | | 0.243* | 0.286** |
| | | (0.026) | (0.026) | | (0.126) | (0.119) | | (0.100) | (0.099) |
| Previous BERT Publications | | 0.000 | -0.002 | | 0.347*** | 0.293*** | | 0.493*** | 0.508*** |
| | | (0.011) | (0.012) | | (0.053) | (0.051) | | (0.042) | (0.042) |
| Field Fixed Effects | YES | YES | YES | YES | YES | YES | YES | YES | YES |
| Publication Time Fixed Effects | | | YES | | | YES | | | YES |
| Constant | 1.359*** | 1.316*** | 1.141*** | 6.342*** | 6.439*** | 4.332*** | 6.324*** | 6.108*** | 5.238*** |
| | (0.040) | (0.059) | (0.087) | (0.188) | (0.272) | (0.366) | (0.153) | (0.215) | (0.304) |
| Adjusted R-squared/Pseudo R2 | 0.008 | 0.023 | 0.026 | 0.045 | 0.093 | 0.198 | 0.025 | 0.123 | 0.143 |
| N | 2633 | 2633 | 2633 | 2629 | 2629 | 2629 | 2629 | 2629 | 2629 |

Note: Standard errors in parentheses. $^+ p < 0.1$, $^* p < 0.05$, $^{**} p < 0.01$, $^{***} p < 0.001$. All tests are two-tailed.

Table 12. Robustness Test 2: Regression Estimates of Generation Effects on Institution Characteristics, Based on a Selective Sample

| | DV: Number of Institutions | | | DV: log (All Paper Counts) | | | DV: Institution Dispersion | | |
|---|---|---|---|---|---|---|---|---|---|
| | Model 1 | Model 2 | Model 3 | Model 4 | Model 5 | Model 6 | Model 7 | Model 8 | Model 9 |
| Paper Generation | 0.009 | 0.013 | 0.013 | 0.026 | 0.017 | 0.010 | 0.006 | 0.008+ | 0.009* |
| | (0.010) | (0.010) | (0.010) | (0.020) | (0.020) | (0.020) | (0.004) | (0.004) | (0.004) |
| Quality: One Year Citations | | 0.012** | 0.012** | | 0.021** | 0.025** | | 0.004* | 0.004* |
| | | (0.004) | (0.004) | | (0.008) | (0.008) | | (0.002) | (0.002) |
| Number of Level 2 Concepts | | 0.001 | -0.000 | | -0.011 | -0.000 | | 0.002 | 0.000 |
| | | (0.007) | (0.008) | | (0.015) | (0.015) | | (0.003) | (0.003) |
| Venue: Conference | | 0.015 | 0.026 | | 0.104 | 0.182 | | 0.005 | 0.006 |
| | | (0.063) | (0.066) | | (0.125) | (0.131) | | (0.025) | (0.027) |
| Venue: Core Collection | | 0.138* | 0.142* | | 0.056 | 0.039 | | 0.037 | 0.040 |
| | | (0.068) | (0.070) | | (0.135) | (0.139) | | (0.027) | (0.028) |
| Venue: Impact Factor | | 0.006 | 0.007 | | 0.032*** | 0.034*** | | 0.002 | 0.003 |
| | | (0.005) | (0.005) | | (0.009) | (0.009) | | (0.002) | (0.002) |
| Any Industrial Affiliation | | 0.283*** | 0.284*** | | -0.956*** | -0.937*** | | 0.131*** | 0.130*** |
| | | (0.038) | (0.039) | | (0.078) | (0.078) | | (0.016) | (0.016) |
| Previous BERT Publications | | 0.010 | 0.009 | | 0.150*** | 0.160*** | | 0.016* | 0.017* |
| | | (0.017) | (0.018) | | (0.035) | (0.036) | | (0.007) | (0.007) |
| Field Fixed Effects | YES | YES | YES | YES | YES | YES | YES | YES | YES |
| Publication Time Fixed Effects | | | YES | | | YES | | | YES |
| Constant | 0.558*** | 0.388*** | 0.373** | 10.124*** | 10.161*** | 9.607*** | 0.254*** | 0.183*** | 0.192*** |
| | (0.061) | (0.091) | (0.131) | (0.127) | (0.183) | (0.255) | (0.025) | (0.037) | (0.052) |
| Adjusted R-squared/Pseudo R2 | 0.004 | 0.016 | 0.018 | 0.014 | 0.089 | 0.100 | 0.010 | 0.049 | 0.051 |
| N | 2633 | 2633 | 2633 | 2283 | 2283 | 2283 | 2283 | 2283 | 2283 |

Note: Standard errors in parentheses. $^{+} p < 0.1$, $^{*} p < 0.05$, $^{**} p < 0.01$, $^{***} p < 0.001$. All tests are two-tailed.

Table 13. Robustness Test 3: Regression Estimates of Generation Effects on International Collaborations, Based on a Selective Sample

| | DV: Number of Countries Involved | | | DV: Country Dispersion | | |
|---|---|---|---|---|---|---|
| | Model 1 | Model 2 | Model 3 | Model 4 | Model 5 | Model 6 |
| Paper Generation | 0.004 | 0.007 | 0.005 | 0.002 | 0.003 | 0.002 |
| | (0.012) | (0.012) | (0.012) | (0.003) | (0.003) | (0.003) |
| Quality: One Year Citations | | 0.009[*] | 0.009[*] | | 0.004[**] | 0.003[**] |
| | | (0.004) | (0.005) | | (0.001) | (0.001) |
| Number of Level 2 Concepts | | 0.005 | 0.006 | | 0.002 | 0.002 |
| | | (0.009) | (0.009) | | (0.002) | (0.002) |
| Venue: Conference | | -0.025 | -0.026 | | -0.015 | -0.017 |
| | | (0.072) | (0.076) | | (0.019) | (0.020) |
| Venue: Core Collection | | 0.020 | 0.011 | | -0.013 | -0.016 |
| | | (0.079) | (0.082) | | (0.021) | (0.021) |
| Venue: Impact Factor | | 0.005 | 0.005 | | 0.002 | 0.002 |
| | | (0.005) | (0.006) | | (0.001) | (0.001) |
| Any Industrial Affiliation | | 0.175[***] | 0.178[***] | | 0.069[***] | 0.070[***] |
| | | (0.046) | (0.047) | | (0.012) | (0.012) |
| Previous BERT Publications | | 0.022 | 0.018 | | 0.016[**] | 0.016[**] |
| | | (0.020) | (0.021) | | (0.005) | (0.005) |
| Field Fixed Effects | YES | YES | YES | YES | YES | YES |
| Publication Time Fixed Effects | | | YES | | | YES |
| Constant | 0.268[***] | 0.170 | 0.186 | 0.107[***] | 0.073[**] | 0.094[*] |
| | (0.072) | (0.106) | (0.152) | (0.019) | (0.028) | (0.039) |
| Adjusted R-squared/Pseudo R2 | 0.001 | 0.006 | 0.007 | 0.012 | 0.040 | 0.038 |
| N | 2633 | 2633 | 2633 | 2283 | 2283 | 2283 |

Table 14. Robustness Test 4: Regression Estimates of Generation Effects on Knowledge Breadth, Based on a Selective Sample

| | DV: Level 0 Concepts | | | DV: Level 1 Concepts | | | DV: Level 2 Concepts | | |
|---|---|---|---|---|---|---|---|---|---|
| | Model 1 | Model 2 | Model 3 | Model 4 | Model 5 | Model 6 | Model 7 | Model 8 | Model 9 |
| Paper Generation | -0.012*** | -0.012*** | -0.011*** | -0.027 | -0.025 | -0.006 | -0.123*** | -0.109*** | -0.058* |
| | (0.003) | (0.003) | (0.003) | (0.019) | (0.019) | (0.019) | (0.026) | (0.026) | (0.026) |
| Quality: One Year Citations | | 0.002 | 0.001 | | 0.013+ | 0.009 | | 0.015 | 0.004 |
| | | (0.001) | (0.001) | | (0.007) | (0.008) | | (0.011) | (0.010) |
| Number of Authors | | 0.003 | 0.003 | | -0.025+ | -0.023 | | -0.115*** | -0.108*** |
| | | (0.003) | (0.003) | | (0.015) | (0.015) | | (0.021) | (0.021) |
| Number of Institutions | | 0.003 | 0.004 | | 0.055* | 0.055* | | 0.055 | 0.047 |
| | | (0.004) | (0.004) | | (0.025) | (0.025) | | (0.036) | (0.035) |
| Venue: Conference | | 0.050* | 0.044* | | 0.445*** | 0.408*** | | 0.323* | 0.265 |
| | | (0.019) | (0.020) | | (0.117) | (0.121) | | (0.164) | (0.167) |
| Venue: Core Collection | | 0.003 | 0.003 | | 0.398** | 0.490*** | | 0.249 | 0.465** |
| | | (0.021) | (0.022) | | (0.128) | (0.130) | | (0.181) | (0.179) |
| Venue: Impact Factor | | -0.002 | -0.002 | | 0.008 | 0.007 | | 0.021 | 0.021 |
| | | (0.002) | (0.002) | | (0.009) | (0.009) | | (0.013) | (0.013) |
| Any Industrial Affiliation | | 0.023+ | 0.021 | | 0.063 | -0.000 | | 0.260* | 0.124 |
| | | (0.013) | (0.014) | | (0.081) | (0.080) | | (0.114) | (0.111) |
| Previous BERT Publications | | 0.003 | 0.001 | | -0.005 | -0.023 | | 0.048 | 0.024 |
| | | (0.006) | (0.006) | | (0.033) | (0.034) | | (0.046) | (0.046) |
| Field Fixed Effects | YES | YES | YES | YES | YES | YES | YES | YES | YES |
| Publication Time Fixed Effects | | | YES | | | YES | | | YES |
| Constant | 2.618*** | 2.561*** | 2.521*** | 5.035*** | 4.597*** | 4.484*** | 5.187*** | 5.143*** | 5.162*** |
| | (0.019) | (0.027) | (0.039) | (0.115) | (0.163) | (0.233) | (0.163) | (0.230) | (0.321) |
| Adjusted R-squared/Pseudo R2 | 0.966 | 0.966 | 0.966 | 0.493 | 0.496 | 0.509 | 0.353 | 0.362 | 0.406 |
| N | 2633 | 2633 | 2633 | 2633 | 2633 | 2633 | 2633 | 2633 | 2633 |

Table 15. Robustness Test 5: Regression Estimates of Generation Effects on Research Impact, Based on a Selective Sample

| | DV: One Year Forward Citations (Paper in and before 2023) | | | DV: Two Year Forward Citations (Paper in and before 2022) | | | DV: Three Year Forward Citations (Paper in and before 2021) | | |
|---|---|---|---|---|---|---|---|---|---|
| | Model 1 | Model 2 | Model 3 | Model 4 | Model 5 | Model 6 | Model 7 | Model 8 | Model 9 |
| Paper Generation | -0.029+ | -0.043** | -0.044** | -0.078*** | -0.096*** | -0.080*** | -0.111*** | -0.121*** | -0.109*** |
| | (0.017) | (0.016) | (0.016) | (0.015) | (0.015) | (0.015) | (0.019) | (0.019) | (0.019) |
| Number of Authors | | 0.086*** | 0.085*** | | 0.066*** | 0.064*** | | 0.066*** | 0.060*** |
| | | (0.013) | (0.012) | | (0.011) | (0.011) | | (0.013) | (0.013) |
| Number of Institutions | | 0.012 | 0.014 | | 0.037+ | 0.050** | | 0.026 | 0.023 |
| | | (0.022) | (0.021) | | (0.020) | (0.019) | | (0.023) | (0.023) |
| Number of Level 2 Concepts | | 0.000 | 0.004 | | 0.022* | 0.018+ | | 0.017 | 0.008 |
| | | (0.012) | (0.011) | | (0.011) | (0.011) | | (0.012) | (0.012) |
| Venue: Conference | | 0.004 | 0.219* | | 0.049 | 0.255* | | 0.098 | 0.342** |
| | | (0.104) | (0.107) | | (0.100) | (0.106) | | (0.116) | (0.130) |
| Venue: Core Collection | | 0.203+ | 0.387*** | | 0.132 | 0.332** | | 0.168 | 0.416** |
| | | (0.112) | (0.115) | | (0.108) | (0.114) | | (0.129) | (0.141) |
| Venue: Impact Factor | | 0.074*** | 0.070*** | | 0.056*** | 0.062*** | | 0.040*** | 0.044*** |
| | | (0.007) | (0.007) | | (0.006) | (0.006) | | (0.008) | (0.008) |
| Any Industrial Affiliation | | 0.291*** | 0.250*** | | 0.226*** | 0.176*** | | 0.198*** | 0.185** |
| | | (0.062) | (0.061) | | (0.054) | (0.053) | | (0.059) | (0.059) |
| Previous BERT Publications | | 0.144*** | 0.143*** | | 0.084*** | 0.101*** | | 0.094** | 0.130*** |
| | | (0.026) | (0.025) | | (0.024) | (0.024) | | (0.033) | (0.035) |
| Field Fixed Effects | YES | YES | YES | YES | YES | YES | YES | YES | YES |
| Publication Time Fixed Effects | | | YES | | | YES | | | YES |
| Constant | 1.251*** | 0.631*** | 1.029*** | 2.266*** | 1.660*** | 1.452*** | 2.484*** | 1.725*** | 1.544*** |
| | (0.100) | (0.157) | (0.192) | (0.082) | (0.144) | (0.174) | (0.434) | (0.449) | (0.456) |

| | | | | | | | | |
|---|---|---|---|---|---|---|---|---|
| Adjusted R-squared/Pseudo R2 | 0.005 | 0.033 | 0.047 | 0.008 | 0.028 | 0.035 | 0.009 | 0.023 | 0.026 |
| N | 2396 | 2396 | 2396 | 1852 | 1852 | 1852 | 1174 | 1174 | 1174 |

*Note: Standard errors in parentheses. $^+ p < 0.1$, $^* p < 0.05$, $^{**} p < 0.01$, $^{***} p < 0.001$. All tests are two-tailed.*

Table 16. Robustness Test 5: Regression Estimates of Ordinal Generation Effects on Research Impact

| | DV: One Year Forward Citations (Paper in and before 2023) | | | DV: Two Year Forward Citations (Paper in and before 2022) | | | DV: Three Year Forward Citations (Paper in and before 2021) | | |
|---|---|---|---|---|---|---|---|---|---|
| | Model 1 | Model 2 | Model 3 | Model 4 | Model 5 | Model 6 | Model 7 | Model 8 | Model 9 |
| Paper Generation: group 2 | -0.001 | -0.092 | -0.057 | -0.101 | -0.175$^{**}$ | -0.124$^*$ | -0.120+ | -0.180$^{**}$ | -0.137$^*$ |
| | (0.068) | (0.064) | (0.063) | (0.058) | (0.055) | (0.055) | (0.064) | (0.061) | (0.061) |
| Paper Generation: group 3 | -0.071 | -0.123+ | -0.142$^*$ | -0.162$^{**}$ | -0.203$^{***}$ | -0.188$^{***}$ | -0.186$^{**}$ | -0.209$^{***}$ | -0.195$^{**}$ |
| | (0.070) | (0.066) | (0.064) | (0.060) | (0.057) | (0.056) | (0.063) | (0.061) | (0.060) |
| Paper Generation: group 4 | -0.038 | -0.095 | -0.073 | -0.222$^{***}$ | -0.285$^{***}$ | -0.217$^{***}$ | -0.332$^{***}$ | -0.384$^{***}$ | -0.326$^{***}$ |
| | (0.066) | (0.062) | (0.063) | (0.058) | (0.055) | (0.056) | (0.066) | (0.064) | (0.065) |
| Controls as in Table 15 | YES | YES | YES | YES | YES | YES | YES | YES | YES |
| Field Fixed Effects | YES | YES | YES | YES | YES | YES | YES | YES | YES |
| Publication Year Fixed Effects | | | YES | | | YES | | | YES |
| Constant | 1.181$^{***}$ | 0.554$^{***}$ | 0.967$^{***}$ | 2.124$^{***}$ | 1.501$^{***}$ | 1.325$^{***}$ | 2.186$^{***}$ | 1.396$^{**}$ | 1.235$^{**}$ |
| | (0.090) | (0.151) | (0.189) | (0.073) | (0.141) | (0.173) | (0.427) | (0.442) | (0.448) |
| Adjusted R-squared/Pseudo R2 | 0.005 | 0.033 | 0.047 | 0.007 | 0.027 | 0.034 | 0.008 | 0.022 | 0.025 |
| N | 2396 | 2396 | 2396 | 1852 | 1852 | 1852 | 1174 | 1174 | 1174 |

*Note: Standard errors in parentheses. $^* p < 0.05$, $^{**} p < 0.01$, $^{***} p < 0.001$. All tests are two-tailed. We cut the generation score into four equal quantile groups, and use the first group as a reference group for the variable.*

Table 17. Robustness Test 6: Regression Estimates of Generation Effects on Research Impact, based on An Alternative Measure

| | DV: One Year Forward Citations (Paper in and before 2023) | | | DV: Two Year Forward Citations (Paper in and before 2022) | | | DV: Three Year Forward Citations (Paper in and before 2021) | | |
|---|---|---|---|---|---|---|---|---|---|
| | Model 1 | Model 2 | Model 3 | Model 4 | Model 5 | Model 6 | Model 7 | Model 8 | Model 9 |
| Paper Generation (Alternative) | -0.488*** | -0.507*** | -0.578*** | -0.499*** | -0.525*** | -0.556*** | -0.532*** | -0.536*** | -0.553*** |
| | (0.034) | (0.034) | (0.039) | (0.025) | (0.026) | (0.031) | (0.028) | (0.028) | (0.033) |
| Number of Authors | | 0.057*** | 0.048*** | | 0.033*** | 0.029*** | | 0.029*** | 0.024*** |
| | | (0.009) | (0.009) | | (0.007) | (0.007) | | (0.007) | (0.007) |
| Number of Institutions | | 0.020 | 0.019 | | 0.029* | 0.032** | | 0.010 | 0.006 |
| | | (0.018) | (0.016) | | (0.013) | (0.012) | | (0.012) | (0.012) |
| Number of Level 2 Concepts | | -0.004 | 0.004 | | 0.010 | 0.009 | | 0.004 | -0.000 |
| | | (0.010) | (0.009) | | (0.007) | (0.007) | | (0.006) | (0.007) |
| Venue: Conference | | -0.227** | 0.034 | | -0.025 | 0.171* | | 0.026 | 0.190* |
| | | (0.075) | (0.080) | | (0.063) | (0.071) | | (0.063) | (0.076) |
| Venue: Core Collection | | 0.040 | 0.207* | | 0.101 | 0.251*** | | 0.082 | 0.224** |
| | | (0.083) | (0.085) | | (0.069) | (0.076) | | (0.068) | (0.078) |
| Venue: Impact Factor | | 0.055*** | 0.048*** | | 0.044*** | 0.041*** | | 0.030*** | 0.027*** |
| | | (0.004) | (0.005) | | (0.003) | (0.003) | | (0.004) | (0.004) |
| Any Industrial Affiliation | | 0.149** | 0.155*** | | 0.115*** | 0.101** | | 0.124*** | 0.120*** |
| | | (0.047) | (0.044) | | (0.035) | (0.034) | | (0.031) | (0.030) |
| Previous BERT Publications | | 0.145*** | 0.132*** | | 0.101*** | 0.096*** | | 0.093*** | 0.091*** |
| | | (0.018) | (0.016) | | (0.015) | (0.015) | | (0.018) | (0.018) |
| Paper Component Fixed Effects | YES | YES | YES | YES | YES | YES | YES | YES | YES |
| Field Fixed Effects | YES | YES | YES | YES | YES | YES | YES | YES | YES |
| Publication Year Fixed Effects | | | YES | | | YES | | | YES |
| Constant | 2.302*** | 2.049*** | 2.047*** | 3.054*** | 2.762*** | 2.631*** | 3.603*** | 3.337*** | 3.253*** |
| | (0.088) | (0.138) | (0.145) | (0.064) | (0.100) | (0.112) | (0.063) | (0.100) | (0.114) |

| | Model 1 | Model 2 | Model 3 | Model 4 | Model 5 | Model 6 | Model 7 | Model 8 | Model 9 |
|---|---|---|---|---|---|---|---|---|---|
| Adjusted R-squared/Pseudo R2 | 0.069 | 0.150 | 0.197 | 0.098 | 0.162 | 0.175 | 0.130 | 0.168 | 0.174 |
| N | 3103 | 3103 | 3103 | 2569 | 2569 | 2569 | 1793 | 1793 | 1793 |

*Note: Standard errors in parentheses. * p < 0.05, ** p < 0.01, *** p < 0.001. All tests are two-tailed.*

Table 18. Regression Estimates of Generation Effects on Author Features: Varying Time Fixed Effects

| | DV: Number of Authors | | | DV: Average Career Age | | | DV: log (Cumulative Citations) | | |
|---|---|---|---|---|---|---|---|---|---|
| | Model 1 | Model 2 | Model 3 | Model 4 | Model 5 | Model 6 | Model 7 | Model 8 | Model 9 |
| **FE Period Time Window** | **1M** | **2M** | **3M** | **1M** | **2M** | **3M** | **1M** | **2M** | **3M** |
| Paper Generation | 0.021** | 0.022*** | 0.021*** | 0.085** | 0.079** | 0.084** | 0.025 | 0.025 | 0.023 |
| | (0.007) | (0.006) | (0.006) | (0.028) | (0.028) | (0.028) | (0.024) | (0.024) | (0.024) |
| Quality: One Year Citations | 0.015*** | 0.014*** | 0.014*** | 0.041*** | 0.037*** | 0.037*** | 0.068*** | 0.065*** | 0.065*** |
| | (0.002) | (0.002) | (0.002) | (0.009) | (0.009) | (0.009) | (0.008) | (0.008) | (0.008) |
| Number of Level 2 Concepts | -0.020*** | -0.020*** | -0.020*** | 0.044** | 0.038* | 0.034* | -0.008 | -0.012 | -0.014 |
| | (0.004) | (0.004) | (0.004) | (0.017) | (0.017) | (0.017) | (0.014) | (0.014) | (0.014) |
| Venue: Conference | 0.142*** | 0.143*** | 0.136*** | 0.178 | 0.194 | 0.172 | 0.486*** | 0.489*** | 0.432*** |
| | (0.037) | (0.036) | (0.035) | (0.155) | (0.150) | (0.145) | (0.133) | (0.127) | (0.124) |
| Venue: Core Collection | 0.143*** | 0.140*** | 0.131*** | 0.319+ | 0.337* | 0.314* | 0.379** | 0.385** | 0.316* |
| | (0.040) | (0.039) | (0.037) | (0.166) | (0.162) | (0.157) | (0.142) | (0.138) | (0.134) |
| Venue: Impact Factor | 0.006* | 0.005* | 0.006* | 0.039*** | 0.039*** | 0.040*** | 0.059*** | 0.058*** | 0.060*** |
| | (0.003) | (0.003) | (0.003) | (0.011) | (0.011) | (0.011) | (0.010) | (0.010) | (0.010) |
| Any Industrial Affiliation | 0.239*** | 0.240*** | 0.240*** | 0.674*** | 0.676*** | 0.670*** | 0.389*** | 0.388*** | 0.383*** |
| | (0.021) | (0.020) | (0.020) | (0.093) | (0.093) | (0.093) | (0.079) | (0.079) | (0.079) |
| Previous BERT Publications | -0.005 | -0.005 | -0.006 | 0.280*** | 0.296*** | 0.297*** | 0.524*** | 0.524*** | 0.526*** |
| | (0.010) | (0.010) | (0.010) | (0.042) | (0.042) | (0.042) | (0.036) | (0.036) | (0.036) |
| Field/Time Fixed Effects | YES | YES | YES | YES | YES | YES | YES | YES | YES |
| Constant | 0.462 | 0.468 | 0.475 | 4.621* | 4.662* | 4.679* | 5.650** | 5.661** | 5.744** |
| | (0.709) | (0.709) | (0.709) | (2.069) | (2.076) | (2.075) | (1.768) | (1.770) | (1.770) |
| Adjusted R-squared/Pseudo R2 | 0.029 | 0.027 | 0.027 | 0.208 | 0.202 | 0.202 | 0.142 | 0.140 | 0.140 |

| N | 4130 | 4130 | 4130 | 4121 | 4121 | 4121 | 4121 | 4121 | 4121 |

*Note: Standard errors in parentheses. $^+ p < 0.1$, $^* p < 0.05$, $^{**} p < 0.01$, $^{***} p < 0.001$. All tests are two-tailed.*

Table 19. Regression Estimates of Generation Effects on Knowledge Breadth: Varying Time Fixed Effects

| | DV: Level 0 Concepts | | | DV: Level 1 Concepts | | | DV: Level 2 Concepts | | |
|---|---|---|---|---|---|---|---|---|---|
| | Model 1 | Model 2 | Model 3 | Model 4 | Model 5 | Model 6 | Model 7 | Model 8 | Model 9 |
| **FE Period Time Window** | **1M** | **2M** | **3M** | **1M** | **2M** | **3M** | **1M** | **2M** | **3M** |
| Paper Generation | -0.008* | -0.008* | -0.008* | -0.004 | -0.007 | -0.005 | -0.048+ | -0.055* | -0.054* |
| | (0.004) | (0.004) | (0.004) | (0.019) | (0.019) | (0.019) | (0.026) | (0.026) | (0.026) |
| Quality: One Year Citations | 0.001 | 0.001 | 0.001 | 0.005 | 0.007 | 0.007 | 0.004 | 0.007 | 0.007 |
| | (0.001) | (0.001) | (0.001) | (0.006) | (0.006) | (0.006) | (0.008) | (0.008) | (0.008) |
| Number of Authors | 0.002 | 0.002 | 0.002 | -0.007 | -0.007 | -0.008 | -0.089*** | -0.093*** | -0.091*** |
| | (0.002) | (0.002) | (0.002) | (0.012) | (0.012) | (0.012) | (0.017) | (0.017) | (0.017) |
| Number of Institutions | 0.003 | 0.003 | 0.004 | 0.036+ | 0.041+ | 0.041* | 0.017 | 0.029 | 0.029 |
| | (0.004) | (0.004) | (0.004) | (0.021) | (0.021) | (0.021) | (0.029) | (0.029) | (0.029) |
| Venue: Conference | 0.050* | 0.046* | 0.043* | 0.327** | 0.307** | 0.326*** | 0.245+ | 0.229+ | 0.250+ |
| | (0.020) | (0.020) | (0.019) | (0.103) | (0.100) | (0.097) | (0.143) | (0.139) | (0.134) |
| Venue: Core Collection | 0.023 | 0.015 | 0.013 | 0.517*** | 0.479*** | 0.507*** | 0.542*** | 0.523*** | 0.560*** |
| | (0.022) | (0.021) | (0.021) | (0.110) | (0.108) | (0.105) | (0.153) | (0.150) | (0.146) |
| Venue: Impact Factor | -0.001 | -0.000 | -0.000 | 0.001 | 0.002 | 0.001 | 0.013 | 0.015 | 0.016 |
| | (0.002) | (0.002) | (0.001) | (0.008) | (0.008) | (0.008) | (0.011) | (0.011) | (0.011) |
| Any Industrial Affiliation | 0.012 | 0.010 | 0.010 | 0.028 | 0.026 | 0.024 | 0.115 | 0.131 | 0.114 |
| | (0.013) | (0.013) | (0.013) | (0.064) | (0.064) | (0.064) | (0.089) | (0.089) | (0.089) |
| Previous BERT Publications | 0.013* | 0.013* | 0.013* | -0.008 | -0.020 | -0.024 | 0.040 | 0.029 | 0.019 |
| | (0.006) | (0.006) | (0.006) | (0.028) | (0.028) | (0.028) | (0.039) | (0.039) | (0.039) |
| Field/Year Fixed Effects | YES | YES | YES | YES | YES | YES | YES | YES | YES |
| Constant | 2.959*** | 2.955*** | 2.960*** | 3.112* | 3.159* | 3.123* | 4.430* | 4.488* | 4.426* |

| | (0.275) | (0.276) | (0.276) | (1.387) | (1.394) | (1.393) | (1.925) | (1.937) | (1.938) |
| Adjusted R-squared/Pseudo R2 | 0.953 | 0.952 | 0.952 | 0.507 | 0.502 | 0.502 | 0.398 | 0.390 | 0.389 |
| N | 4130 | 4130 | 4130 | 4130 | 4130 | 4130 | 4130 | 4130 | 4130 |

*Note: Standard errors in parentheses. $^+ p < 0.1, ^* p < 0.05, ^{**} p < 0.01, ^{***} p < 0.001$. All tests are two-tailed.*

Table 20. Regression Estimates of Generation Effects on Research Impact: Varying Time Fixed Effects

| | DV: One Year Forward Citations (Paper in and before 2023) | | | DV: Two Year Forward Citations (Paper in and before 2022) | | | DV: Three Year Forward Citations (Paper in and before 2021) | | |
|---|---|---|---|---|---|---|---|---|---|
| | Model 1 | Model 2 | Model 3 | Model 4 | Model 5 | Model 6 | Model 7 | Model 8 | Model 9 |
| **FE Period Time Window** | **1M** | **2M** | **3M** | **1M** | **2M** | **3M** | **1M** | **2M** | **3M** |
| Paper Generation | -0.027+ | -0.027+ | -0.027+ | -0.069*** | -0.068*** | -0.069*** | -0.097*** | -0.097*** | -0.099*** |
| | (0.016) | (0.016) | (0.016) | (0.014) | (0.014) | (0.015) | (0.018) | (0.018) | (0.018) |
| Number of Authors | 0.069*** | 0.069*** | 0.069*** | 0.051*** | 0.053*** | 0.052*** | 0.043*** | 0.044*** | 0.043*** |
| | (0.010) | (0.010) | (0.010) | (0.009) | (0.009) | (0.009) | (0.010) | (0.010) | (0.010) |
| Number of Institutions | 0.015 | 0.017 | 0.018 | 0.044** | 0.041** | 0.044** | 0.024 | 0.021 | 0.024 |
| | (0.017) | (0.017) | (0.017) | (0.015) | (0.015) | (0.015) | (0.017) | (0.017) | (0.017) |
| Number of Level 2 Concepts | 0.002 | 0.004 | 0.004 | 0.017* | 0.017* | 0.017* | 0.008 | 0.010 | 0.010 |
| | (0.009) | (0.009) | (0.009) | (0.008) | (0.008) | (0.008) | (0.009) | (0.009) | (0.009) |
| Venue: Conference | 0.382*** | 0.332*** | 0.306*** | 0.572*** | 0.473*** | 0.407*** | 0.661*** | 0.505*** | 0.386*** |
| | (0.095) | (0.092) | (0.086) | (0.098) | (0.092) | (0.083) | (0.120) | (0.109) | (0.099) |
| Venue: Core Collection | 0.505*** | 0.477*** | 0.447*** | 0.652*** | 0.561*** | 0.493*** | 0.746*** | 0.603*** | 0.484*** |
| | (0.100) | (0.098) | (0.091) | (0.103) | (0.098) | (0.089) | (0.126) | (0.114) | (0.105) |
| Venue: Impact Factor | 0.066*** | 0.069*** | 0.069*** | 0.061*** | 0.061*** | 0.061*** | 0.043*** | 0.046*** | 0.046*** |
| | (0.006) | (0.006) | (0.006) | (0.005) | (0.005) | (0.005) | (0.006) | (0.006) | (0.006) |
| Any Industrial Affiliation | 0.167*** | 0.178*** | 0.183*** | 0.166*** | 0.166*** | 0.174*** | 0.172*** | 0.173*** | 0.178*** |
| | (0.048) | (0.048) | (0.048) | (0.041) | (0.041) | (0.041) | (0.044) | (0.044) | (0.044) |
| Previous BERT Publications | 0.177*** | 0.171*** | 0.172*** | 0.143*** | 0.132*** | 0.134*** | 0.169*** | 0.160*** | 0.162*** |
| | (0.021) | (0.021) | (0.021) | (0.020) | (0.020) | (0.020) | (0.026) | (0.026) | (0.026) |
| Field/Year Fixed Effects | YES | YES | YES | YES | YES | YES | YES | YES | YES |
| Constant | -0.012 | -0.015 | 0.021 | 1.579* | 1.677* | 1.754* | 1.637 | 1.852* | 1.926* |

| | (1.128) | (1.137) | (1.138) | (0.793) | (0.798) | (0.800) | (0.837) | (0.828) | (0.831) |
| Adjusted R-squared/Pseudo R2 | 0.052 | 0.049 | 0.049 | 0.036 | 0.034 | 0.033 | 0.028 | 0.026 | 0.025 |
| N | 3835 | 3835 | 3835 | 3109 | 3109 | 3109 | 2159 | 2159 | 2159 |

*Note: Standard errors in parentheses. $^+ p < 0.1$, $^* p < 0.05$, $^{**} p < 0.01$, $^{***} p < 0.001$. All tests are two-tailed.*

Table 21. Regression Estimates of Generation Effects on Research Impact: Citations (Deleting Self-Citations)

| | DV: One Year Forward Citations (Paper in and before 2023) | | | DV: Two Year Forward Citations (Paper in and before 2022) | | | DV: Three Year Forward Citations (Paper in and before 2021) | | |
|---|---|---|---|---|---|---|---|---|---|
| | Model 1 | Model 2 | Model 3 | Model 4 | Model 5 | Model 6 | Model 7 | Model 8 | Model 9 |
| Paper Generation | -0.020* | -0.051** | -0.046** | -0.020* | -0.101*** | -0.084*** | -0.026** | -0.130*** | -0.114*** |
| | (0.010) | (0.018) | (0.018) | (0.009) | (0.016) | (0.016) | (0.010) | (0.020) | (0.020) |
| Number of Authors | | 0.065*** | 0.066*** | | 0.048*** | 0.048*** | | 0.045*** | 0.040*** |
| | | (0.011) | (0.011) | | (0.009) | (0.009) | | (0.010) | (0.010) |
| Number of Institutions | | -0.012 | -0.010 | | 0.021 | 0.026 | | 0.004 | 0.002 |
| | | (0.019) | (0.019) | | (0.017) | (0.016) | | (0.018) | (0.018) |
| Number of Level 2 Concepts | | 0.007 | 0.007 | | 0.022* | 0.020* | | 0.019* | 0.011 |
| | | (0.010) | (0.010) | | (0.009) | (0.009) | | (0.009) | (0.009) |
| Venue: Conference | | 0.068 | 0.221* | | 0.168* | 0.356*** | | 0.169+ | 0.371*** |
| | | (0.087) | (0.092) | | (0.078) | (0.087) | | (0.087) | (0.101) |
| Venue: Core Collection | | 0.176+ | 0.363*** | | 0.237** | 0.439*** | | 0.242* | 0.476*** |
| | | (0.095) | (0.098) | | (0.085) | (0.093) | | (0.096) | (0.107) |
| Venue: Impact Factor | | 0.075*** | 0.074*** | | 0.056*** | 0.063*** | | 0.037*** | 0.043*** |
| | | (0.007) | (0.007) | | (0.005) | (0.006) | | (0.007) | (0.007) |
| Any Industrial Affiliation | | 0.259*** | 0.212*** | | 0.223*** | 0.184*** | | 0.200*** | 0.181*** |
| | | (0.055) | (0.053) | | (0.045) | (0.045) | | (0.048) | (0.047) |
| Previous BERT Publications | | 0.117*** | 0.148*** | | 0.074*** | 0.112*** | | 0.085** | 0.151*** |
| | | (0.023) | (0.023) | | (0.021) | (0.022) | | (0.026) | (0.028) |
| Field Fixed Effects | YES | YES | YES | YES | YES | YES | YES | YES | YES |
| Publication Time Fixed Effects | | | YES | | | YES | | | YES |
| Constant | 1.137*** | 0.678*** | 1.312*** | 1.942*** | 1.561*** | 1.582*** | 1.388** | 1.195** | 1.161* |

| | | | | | | | | |
|---|---|---|---|---|---|---|---|---|
| (0.080) | (0.142) | (0.159) | (0.064) | (0.123) | (0.140) | (0.433) | (0.449) | (0.451) |
| Adjusted R-squared/Pseudo R2 | 0.004 | 0.024 | 0.041 | 0.005 | 0.021 | 0.028 | 0.004 | 0.016 | 0.022 |
| N | 3835 | 3835 | 3835 | 3109 | 3109 | 3109 | 2159 | 2159 | 2159 |

*Note: Standard errors in parentheses. $^+ p < 0.1$, $^* p < 0.05$, $^{**} p < 0.01$, $^{***} p < 0.001$. All tests are two-tailed.*

Table 22. Regression Estimates of Generation Effects on Knowledge Breadth

| | DV: Level 0 Concepts | | | DV: Level 1 Concepts | | | DV: Level 2 Concepts | | |
|---|---|---|---|---|---|---|---|---|---|
| | Model 1 | Model 2 | Model 3 | Model 4 | Model 5 | Model 6 | Model 7 | Model 8 | Model 9 |
| **Concept Inclusion Thresholds** | **0.05** | **0.1** | **0.2** | **0.05** | **0.1** | **0.2** | **0.05** | **0.1** | **0.2** |
| Paper Generation | 0.008 | 0.003 | 0.002 | 0.002 | 0.012 | -0.005 | -0.074** | -0.072** | -0.069** |
| | (0.008) | (0.008) | (0.005) | (0.015) | (0.014) | (0.013) | (0.026) | (0.026) | (0.026) |
| Quality: One Year Citations | -0.004 | -0.002 | -0.005** | 0.005 | 0.006 | 0.006 | 0.015+ | 0.016+ | 0.015+ |
| | (0.003) | (0.002) | (0.001) | (0.005) | (0.004) | (0.004) | (0.008) | (0.008) | (0.008) |
| Number of Authors | 0.017** | 0.021*** | 0.012*** | 0.011 | 0.011 | 0.007 | -0.087*** | -0.085*** | -0.087*** |
| | (0.006) | (0.005) | (0.003) | (0.010) | (0.009) | (0.008) | (0.017) | (0.017) | (0.017) |
| Number of Institutions | -0.013 | 0.006 | 0.015** | 0.013 | 0.023 | 0.018 | 0.036 | 0.035 | 0.039 |
| | (0.009) | (0.009) | (0.005) | (0.017) | (0.016) | (0.014) | (0.029) | (0.029) | (0.029) |
| Venue: Conference | -0.212*** | -0.224*** | -0.153*** | 0.001 | -0.011 | 0.115+ | 0.322* | 0.314* | 0.324** |
| | (0.041) | (0.038) | (0.023) | (0.071) | (0.067) | (0.061) | (0.126) | (0.126) | (0.125) |
| Venue: Core Collection | -0.244*** | -0.186*** | -0.099*** | 0.077 | 0.090 | 0.167* | 0.350* | 0.354* | 0.355* |
| | (0.045) | (0.042) | (0.026) | (0.080) | (0.075) | (0.068) | (0.141) | (0.141) | (0.140) |
| Venue: Impact Factor | -0.006+ | -0.002 | 0.002 | -0.005 | -0.006 | -0.005 | 0.013 | 0.013 | 0.011 |
| | (0.003) | (0.003) | (0.002) | (0.006) | (0.006) | (0.005) | (0.010) | (0.010) | (0.010) |
| Any Industrial Affiliation | -0.051+ | -0.028 | -0.027 | -0.043 | -0.054 | 0.014 | 0.260** | 0.250** | 0.266** |
| | (0.029) | (0.027) | (0.016) | (0.051) | (0.048) | (0.044) | (0.090) | (0.090) | (0.090) |
| Previous BERT Publications | 0.018 | 0.005 | 0.005 | -0.017 | -0.021 | -0.021 | 0.044 | 0.042 | 0.042 |
| | (0.012) | (0.012) | (0.007) | (0.022) | (0.021) | (0.019) | (0.038) | (0.038) | (0.038) |
| Field Fixed Effects | YES | YES | YES | YES | YES | YES | YES | YES | YES |
| Publication Time Fixed Effects | YES | YES | YES | YES | YES | YES | YES | YES | YES |
| Constant | 1.357*** | 1.266*** | 1.151*** | 3.245*** | 3.030*** | 2.562*** | 4.690*** | 4.619*** | 4.468*** |

|  | (0.064) | (0.059) | (0.036) | (0.111) | (0.106) | (0.095) | (0.197) | (0.197) | (0.195) |
|---|---|---|---|---|---|---|---|---|---|
| Adjusted R-squared/Pseudo R2 | 0.513 | 0.362 | 0.256 | 0.097 | 0.080 | 0.049 | 0.307 | 0.305 | 0.305 |
| N | 4130 | 4130 | 4130 | 4130 | 4130 | 4130 | 4130 | 4130 | 4130 |

*Note: Standard errors in parentheses.* [+] *p < 0.1,* [*] *p < 0.05,* [**] *p < 0.01,* [***] *p < 0.001. All tests are two-tailed.*

Table 23. Regression Estimates of Generation Effects on Industrial Participation

| | DV: Industrial Participation (Full Sample) | | DV: Industrial Participation (Selective Sample) | |
|---|---|---|---|---|
| | Model 1 | Model 2 | Model 3 | Model 4 |
| **Paper Generation** | 0.005 | -0.061 | -0.017 | -0.046 |
| | (0.022) | (0.043) | (0.041) | (0.043) |
| **Publication Year** | -0.173*** | -0.190*** | -0.235*** | -0.244*** |
| | (0.035) | (0.040) | (0.045) | (0.052) |
| Number of Authors | | 0.227*** | | 0.201*** |
| | | (0.024) | | (0.030) |
| Number of Institutions | | 0.249*** | | 0.264*** |
| | | (0.038) | | (0.047) |
| Quality: One Year Citations | | 0.048*** | | 0.060*** |
| | | (0.011) | | (0.014) |
| Number of Level 2 Concepts | | 0.057* | | 0.059+ |
| | | (0.024) | | (0.030) |
| Venue: Conference | | 0.188 | | 0.205 |
| | | (0.200) | | (0.262) |
| Venue: Core Collection | | -0.688** | | -0.728* |
| | | (0.233) | | (0.303) |
| Venue: Impact Factor | | 0.027* | | 0.034* |
| | | (0.014) | | (0.017) |
| Previous BERT Publications | | 0.143** | | 0.106 |
| | | (0.053) | | (0.066) |
| Field Fixed Effects | YES | YES | YES | YES |
| Constant | 348.573*** | 381.554*** | 473.906*** | 489.282*** |
| | (71.465) | (81.381) | (91.904) | (104.688) |
| Adjusted R-squared/Pseudo R2 | 0.019 | 0.104 | 0.031 | 0.115 |
| N | 4128 | 4128 | 2631 | 2631 |

*Note: Standard errors in parentheses. $^+ p < 0.1,$ $^* p < 0.05,$ $^{**} p < 0.01,$ $^{***} p < 0.001$. All tests are two-tailed.*